\documentclass[journal,draftcls,onecolumn,twoside]{IEEEtran}\IEEEoverridecommandlockouts
\usepackage{amsmath, amsfonts, amssymb, amsthm}
\usepackage{algorithmic}
\usepackage{algorithm}
\usepackage{array}
\usepackage[caption=false,font=normalsize,labelfont=sf,textfont=sf]{subfig}
\usepackage{textcomp}
\usepackage{stfloats}
\usepackage{color}
\usepackage{url}
\usepackage{verbatim}
\usepackage{graphicx}
\usepackage{cite}
\usepackage{tikz}
\usetikzlibrary{fit}
\usepackage{tabularx}
\usepackage{multirow}
\usepackage{booktabs}
\usepackage{changepage}
\usepackage{makecell}
\newcommand{\RNum}[1]{\uppercase\expandafter{\romannumeral #1\relax}}

\theoremstyle{definition} \newtheorem{theorem}{Theorem}
\theoremstyle{definition} 
\theoremstyle{definition} \newtheorem{proposition}[theorem]{Proposition}
\theoremstyle{definition} 
\theoremstyle{definition} \newtheorem{lemma}[theorem]{Lemma}
\theoremstyle{definition} \newtheorem{example}{Example}
\theoremstyle{definition} 
\theoremstyle{definition} \newtheorem{G-construction}{Generic Construction} 

\theoremstyle{definition} 
\theoremstyle{definition} \newtheorem*{notation}{Notation}

\newcommand{\scalematrix}[2][0.9]{\scalebox{#1}{$\displaystyle #2$}}

{\end{list}}

\begin{document}

\title{Generic Construction of Optimal-Access Binary MDS Array Codes with Smaller Sub-packetization}
\author{\IEEEauthorblockN{Lan~Ma{$^1$},~~Qifu~Tyler~Sun{$^{1*}$},~Shaoteng~Liu{$^{2}$}, ~Liyang~Zhou{$^{2}$}} \\
\IEEEauthorblockA{
{$^1$}School of Computer \& Communication Engineering, University of Science and Technology Beijing, Beijing, China \\
{$^2$}Network Technology Lab, Huawei Technologies Co., Ltd., Shenzhen, China
}

\thanks{$^*~$ Q. T. Sun (Email: qfsun@ustb.edu.cn) is the corresponding author.}
}

\maketitle
\pagestyle{empty}  
\thispagestyle{empty} 
\sloppy

\begin{abstract}
A $(k+r, k, l)$ binary array code of length $k+r$, dimension $k$, and sub-packetization $l$ is composed of $l \times (k+r)$ matrices over $\mathbb{F}_2$, with every column of the matrix stored on a separate node in the distributed storage system and viewed as a coordinate of the codeword. %
It is said to be maximum distance separable (MDS) if any $k$ out of $k+r$ coordinates suffice to reconstruct the whole codeword. %
The repair problem of binary MDS array codes has drawn much attention, particularly for single-node failures. %
In this paper, given an arbitrary binary MDS array code with sub-packetization $m$ as the base code, we propose two generic approaches (Generic Construction \ref{G-construction1} and \ref{cons:G-cons-optrepair}) for constructing binary MDS array codes with optimal access bandwidth or optimal repair bandwidth for single-node failures. %
For every $s \leq r$, a $(k+r, k, ms^{\lceil \frac{k+r}{s}\rceil})$ code $\mathcal{C}_1$ with optimal access bandwidth can be constructed by Generic Construction \ref{G-construction1}. %
Repairing a failed node of $\mathcal{C}_1$ requires connecting to $d = k+s-1$ helper nodes, in which $s-1$ helper nodes are designated and $k$ are free to select. %
The architectural differences of the parity-check matrices between $\mathcal{C}_1$ and previously proposed binary MDS array codes with optimal access bandwidth and with the smallest sub-packetization are illustrated. %
Moreover, $\mathcal{C}_1$ generally achieves smaller sub-packetization and provides greater flexibility in the selection of its coefficient matrices. %
For even $r \geq 4$ and $s = \frac{r}{2}$ such that $s+1$ divides $k+r$, a $(k+r, k, ms^{\frac{k+r}{s+1}})$ code $\mathcal{C}_2$ with optimal repair bandwidth can be constructed by Generic Construction \ref{cons:G-cons-optrepair}, with $\frac{s}{s+1}(k+r)$ out of $k+r$ nodes having the optimal access property. %
In order to achieve the optimal repair bandwidth, the average number of data bits accessed for repairing any single-node failure of $\mathcal{C}_2$ is $\frac{2s}{s+1}\frac{dl}{d-k+1}$, less than twice the optimal one.
To the best of our knowledge, $\mathcal{C}_2$ possesses the smallest sub-packetization among existing binary MDS array codes with optimal repair bandwidth known to date, and among existing binary MDS array codes with optimal repair but not optimal access bandwidth, $\mathcal{C}_2$ has the smallest average number of data bits accessed for repairing any single-node failure. %
\end{abstract}

\begin{IEEEkeywords}
Distributed storage system, repair bandwidth, optimal repair, optimal access, binary MDS array codes.
\end{IEEEkeywords}

\section{Introduction}

Distributed storage systems (DSS) employ redundancy mechanisms to ensure fault tolerance and data reliability. %
Traditional DSS implementations, including the Google File System \cite{Ghem-Goggle} and Hadoop Distributed File System (HDFS) \cite{Shva-HDFS}, predominantly utilize data replication as their primary redundancy strategy. %
However, replication becomes prohibitively expensive as the amount of data grows exponentially. %
In contrast, erasure coding is another redundancy technique that offers superior reliability than replication at the same storage overhead \cite{Repli-erasure}, and has been widely adopted in modern systems such as those built upon CEPH \cite{Weil-CEPH}, which often leverage libraries like Intel Intelligent Storage Acceleration Library (ISA-L) \cite{ISAL} for performance acceleration. %
Among erasure codes, \emph{maximum distance separable (MDS)} codes are particularly significant, providing optimal reliability for a given redundancy level. %
Specifically, a $(k+r,k)$ MDS code over the finite field $\mathbb{F}_q$ encodes $k$ information symbols into $k+r$ encoded symbols of the same size, such that all $k$ information symbols can be reconstructed from any $k$ out of $k+r$ symbols. %
Here, $r$ denotes the number of parity symbols. %

DSS frequently experience node failures, with single-node failures being the most common scenario in production environments due to hardware degradation, network issues, or software faults \cite{single-fail}. %
To maintain system reliability, a self-sustaining DDS must effectively repair such failures. %
In systems utilizing MDS codes, a failed node can be repaired by downloading $k$ symbols from any $k$ surviving nodes. %
However, such a repair strategy is non-optimal for single-node failure recovery in the sense that it incurs a communication overhead of $k$ times the volume of the lost data. %
Consequently, it is critical to design an efficient repair strategy for single-node failures while also preserving the system's inherent fault tolerance for multiple-node failures. %
With respect to a given $k \leq d \leq k+r-1$, define the amount of symbols downloaded from $d$ helper nodes for repairing a single-node failure as the \emph{repair bandwidth}. %
The seminal work \cite{DSS-repair} formulated the node repair problem and established  a lower bound on the repair bandwidth for MDS codes, which motivated a line of research into MDS codes with optimal repair bandwidth. %
While minimizing network transfer remains important, reducing disk I/O overhead is often more critical in practical applications, making the design of MDS codes with optimal access bandwidth highly desirable. %
Various constructions of MDS codes with optimal repair or access bandwidth over non-binary finite fields have been proposed in \cite{MSR-5, MSR-6, MSR-7, MSR-8, Ye-TIT17-opt-repair, Ye-TIT17-opt-access, Li-TIT24-MSR-best-paper, Hou-arxiv-transform} and the references therein. %

A special class of MDS codes called \emph{binary MDS array codes}, which have low computational complexity since the encoding and decoding procedures only involve XOR operations, has attracted much attention in recent years. %
A number of binary MDS array codes have been extensively studied in the literature, such as EVENODD code \cite{Blaum-Evenodd-TComP95} and its generalizations \cite{Blaum-TIT96-MDS-Generalized_EvenOdd} \cite{Blaum-01-Generalized-EvenOdd-Cahpter}, RDP code \cite{RDP_2004} and its generalizations \cite{Blaum-ISIT06-generalized-RDP}, STAR codes \cite{STAR}, Rabin-like codes \cite{Rabin-like}, and various other code constructions presented in \cite{Lv_2023_TIT, Yu_2024_TCom, Jin-Arxiv-circular-shift-based-LNC, Zhai-TCOM25-MDS-array-codes, Lan-array-code}. %
In this paper, we also focus on constructing binary MDS array codes. %
A $(k+r, k, l)$ binary MDS array code is composed of $l \times (k+r)$ matrices over $\mathbb{F}_2$, such that any $k$ out of $k+r$ columns of the matrix suffice to reconstruct the remaining $r$ columns. %
Each column of the matrix is a codeword coordinate. %
The parameter $l$, representing the number of bits per column, is termed as the \emph{sub-packetization level}.
Following the literature on distributed storage, we assume that a binary MDS array code of length $k+r$ composed of $k$ information coordinates and $r$ parity coordinates is distributed across $k+r$ distinct storage nodes, so we use the terms ``coordinate'' and ``node'' interchangeably in this paper. %

Extensive research has addressed the repair problem for binary MDS array codes \cite{butterfly1, butterfly2, MDR-1, MDR-2, Hou-asympto1, Hou-asympto2, Hou-asympto3, Lei-TCOM24-Binary-optimal, Hou-20TIT-transform, Tang-22TCOM-transform, Hou-25TIT-Generic-optimal-repair}. %
Early work include ButterFly codes \cite{butterfly1} \cite{butterfly2} and MDR codes \cite{MDR-1} \cite{MDR-2}, which can achieve optimal repair bandwidth for any information node failure but are limited to $r = 2$ parity nodes. %
Binary MDS array codes with $r > 2$ constructed in \cite{Hou-asympto1} \cite{Hou-asympto2} have asymptotically optimal repair bandwidth for any information node failure. %
Two new classes of binary MDS array codes were constructed in \cite{Hou-asympto3} so that the new codes not only asymptotically achieve optimal repair bandwidth for any information node failure but also exactly achieve optimal repair bandwidth for any parity node failure. %
By stacking multiple instances of Blaum-Roth codes over the polynomial ring $\mathcal{R} = \mathbb{F}_2[x]/(1+x+x^2+\cdots+x^{L-1})$, the work in \cite{Lei-TCOM24-Binary-optimal} constructed two classes of $(k+r, k, (L-1)s^{k+r})$ binary MDS array codes with $s \leq r$ so that under different settings of $L$, one class of codes exactly achieves optimal repair bandwidth for any single-node failure and the other class exactly achieves optimal access bandwidth for any single-node failure, both using $d = k+s-1$ helper nodes. %
Based on the concept of an $s$-pairwise set of binary MDS array codes build upon an arbitrarily given binary MDS array code, the work in \cite{Hou-25TIT-Generic-optimal-repair} proposed a generic construction for constructing $(k+r, k, ms^{k+r})$ binary MDS array codes with optimal repair bandwidth for any single-node failure using $d = k+s-1$ helper nodes. %
Here, $m$ is the sub-packetization of each code in the $s$-pairwise set. %
When the $s$-pairwise set is built upon Blaum-Roth codes, the resulting code construction yields the same class of codes with optimal repair bandwidth as those obtained in \cite{Lei-TCOM24-Binary-optimal}. %
Building on the $(k+r, k, m)$ EVEODD codes \cite{Blaum-Evenodd-TComP95, Blaum-TIT96-MDS-Generalized_EvenOdd} (where $m+1$ is prime and $k \leq m$), the work in \cite{Hou-20TIT-transform} proposed a $(k+r, k, ms^{\lceil \frac{k}{s}\rceil + \lceil\frac{r}{s} \rceil})$ binary MDS array code with optimal access bandwidth. %
The repair of a failed node requires connecting to $d = k+s-1$ helper nodes, in which at least $s-1$ nodes are designated. %
Another generic transformation proposed in \cite{Tang-22TCOM-transform} converts any $(k+r, k, m)$ binary MDS array code with even $m$ into a new $(k+r, k, mr^{\lceil \frac{k+r}{r} \rceil})$ code with optimal access bandwidth using $d = k+r-1$ helper nodes. %
It was remarked in \cite{Tang-22TCOM-transform} that this construction can potentially be extended to $s \leq r$. Based on this extension, the resulting sub-packetization for the general case is $ms^{\lceil \frac{k}{s}\rceil + \lceil\frac{r}{s} \rceil}$, same as the sub-packetization of the code constructed in \cite{Hou-20TIT-transform}.  %
To the best of our knowledge, the smallest sub-packetization for existing binary MDS array codes with optimal access bandwidth is $ms^{\lceil \frac{k}{s}\rceil + \lceil\frac{r}{s} \rceil}$. %

In this work, given an arbitrary binary MDS array code with sub-packetization $m$ as the base code, we introduce two new generic approaches to construct $(k+r, k, l)$ binary MDS array codes with optimal access bandwidth or optimal repair bandwidth for any single-node failure. %
The main contributions are summarized as follows. %
\begin{itemize}
\item For every $s \leq r$, the first generic approach, termed as Generic Construction \ref{G-construction1}, can construct a $(k+r, k, ms^{\lceil \frac{k+r}{s} \rceil}$) binary MDS array code $\mathcal{C}_1$ with optimal access bandwidth. %
    The repair of a failed node of $\mathcal{C}_1$ requires connecting to $d = k+s-1$ helper nodes, in which exactly $s-1$ helper nodes are designated and $k$ are free to select. %
    As the structure of $\mathcal{C}_1$ is described from the perspective of its parity-check matrix, we also interpret the codes in \cite{Hou-20TIT-transform} and \cite{Tang-22TCOM-transform} from the same perspective, so that the structural difference of $\mathcal{C}_1$ from the aforementioned codes become clear. %
    Furthermore, $\mathcal{C}_1$ provides greater flexibility in the selection of its coefficient matrices, in contrast to the fixed coefficient matrices used in the aforementioned codes. %
\item Compared with the sub-packetization $ms^{\lceil \frac{k}{s} \rceil + \lceil \frac{r}{s} \rceil}$ of the code in \cite{Hou-20TIT-transform}, $\mathcal{C}_1$ generally achieves smaller sub-packetization $ms^{\lceil \frac{k+r}{s} \rceil}$. %
    Although \cite{Tang-22TCOM-transform} claimed that its construction can be extended to $s \leq r$, the resulting sub-packetization for  the general case is $ms^{\lceil \frac{k}{s} \rceil+\lceil \frac{r}{s}\rceil}$, which is generally larger than the sub-packetization $ms^{\lceil \frac{k+r}{s} \rceil}$ of $\mathcal{C}_1$. %
    To the best of our knowledge, $\mathcal{C}_1$ generally possesses the smallest sub-packetization among the existing binary MDS array codes with optimal access bandwidth known to date. %
\item For even $r \geq 4$ and $s = \frac{r}{2}$ such that $s+1$ divides $k+r$, another generic approach, termed as Generic Construction \ref{cons:G-cons-optrepair}, can construct a $(k+r, k, ms^{\frac{k+r}{s+1}}$) binary MDS array code $\mathcal{C}_2$ with optimal repair bandwidth using $d = k+s-1$ helper nodes. %
    To the best of our knowledge, compared to existing binary MDS array codes with optimal repair bandwidth, code $\mathcal{C}_1$ achieves reduced sub-packetization $ms^{\lceil \frac{k+r}{s} \rceil}$, while code $\mathcal{C}_2$ attains even lower sub-packetization $ms^{\frac{k+r}{s+1}}$, a value that is the smallest for this class of codes known to date. %
\item For the $(k+r, k, l=s^{k+r})$ codes in \cite{Lei-TCOM24-Binary-optimal} and \cite{Hou-25TIT-Generic-optimal-repair}, in order to achieve optimal repair bandwidth $\frac{dl}{s}$, the amount of the data bits accessed for repairing any single-node failure is $dl$, which is exactly $s$ times the optimal one. %
    In contrast, for $\mathcal{C}_2$ with sub-packetization $l = s^{\frac{k+r}{s+1}}$, repairing each of the $\frac{s}{s+1}(k+r)$ nodes requires  accessing $\frac{l}{s}$ bits from each helper node, while repairing each of the remaining $\frac{1}{s+1}(k+r)$ nodes needs to access $l$ bits from each helper node. %
    Thus, the average number of data bits accessed for repairing any single-node failure of $\mathcal{C}_2$ is $\frac{2s}{s+1}\frac{dl}{d-k+1}$, less than twice the optimal one. %
    To the best of our knowledge, $\mathcal{C}_2$ has the smallest average number of data bits accessed for repairing any single-node failure among existing binary MDS array codes with optimal repair but not optimal access bandwidth. %
\end{itemize}

The remainder of this paper is organized as follows. %
Sec. \ref{sec:Pre} gives some necessary preliminaries. %
Sec. \ref{sec:G-cons-1} presents Generic Construction \ref{G-construction1} for constructing binary MDS array codes with optimal access bandwidth, along with proofs of the MDS and the optimal access properties. %
Sec. \ref{sec:G-cons-optrepair} introduces Generic Construction \ref{cons:G-cons-optrepair} for constructing binary MDS array codes with optimal repair bandwidth, and establishes its key properties. %
Finally, Sec. \ref{sec:Consclusion} draws the conclusion.  %

\begin{notation}
For $0 \leq a \leq s^t-1$, denote by $\mathbf{s}_a =  [a_{t-1} ~ a_{t-2} ~ \cdots ~ a_0 ]$ the $s$-ary expression of $a$, \emph{i.e.},
\begin{equation*}
a = \sum\nolimits_{i=0}^{t-1} a_i s^{i}, 
\end{equation*}
where $0 \leq a_{i} \leq s-1$. %
For $0 \leq a \leq s^t-1$, denote by $a(v,u)$, $0 \leq v \leq t-1$ and $0 \leq u \leq s-1$, the integer smaller than $s^t$ with $\mathbf{s}_{a(v,u)} = [a_{t-1}~a_{t-2}~\dots~a_{v+1}~u~a_{v-1}~\dots~a_0]$. %
The Kronecker product is denoted by $\otimes$. %
The transpose of a matrix $\mathbf{A}$ is denoted as $\mathbf{A}^{\rm T}$. %
Let $\mathbf{I}_t$ denote the $t \times t$ identity matrix. %
In addition, $\mathbf{0}$ and $\mathbf{1}$ respectively represent an all-zero and all-one matrix, whose size, if not explicitly explained, can be inferred in the context. %
$[\mathbf{A}_{ij}]_{0 \leq i \leq i'-1, 0 \leq j \leq j'-1}$ refers to the $i' \times j'$ block matrix, in which every block $\mathbf{A}_{ij}$ is the block entry with row and column respectively indexed by $i$ and $j$. %
Throughout this paper, unless otherwise specified, all indices for rows and columns in a matrix start from zero. 
\end{notation}

\section{Preliminaries}
\label{sec:Pre}

\subsection{Description of binary array codes}
A $(k+r, k, l)$ binary array code, denoted by $\mathcal{C}$, can be viewed as a set of matrices of size $l \times (k+r)$ over $\mathbb{F}_2$. %
Let $[\mathbf{c}_0, \mathbf{c}_1, \cdots, \mathbf{c}_{k+r-1}]$ denote a codeword of $\mathcal{C}$, where each coordinate $\mathbf{c}_j$ is a column vector of $l$ bits. %
These $\mathbf{c}_j$ are referred to as the nodes of $\mathcal{C}$. %
We define the array code $\mathcal{C}$ from the perspective of the parity-check matrix $\mathbf{H}$, which is an $r \times (k+r)$ block matrix in the form \begin{equation}
\label{eqn:PC-matrix-H}
\mathbf{H} = [\mathbf{H}_{i, j}]_{0 \leq i \leq r-1, 0 \leq j \leq k+r-1}, %
\end{equation}
where every $\mathbf{H}_{i,j}$ is an $l\times l$ matrix over $\mathbb{F}_2$. %
Specifically, given a parity-check matrix $\mathbf{H}$, the array code $\mathcal{C}$ is defined as
\begin{equation}
\label{eqn:def-C}
\mathcal{C} = \{ [\mathbf{c}_0, \mathbf{c}_1, \cdots, \mathbf{c}_{k+r-1}]: \sum\nolimits_{j = 0}^{k+r-1} \mathbf{H}_{i, j}\mathbf{c}_j = \mathbf{0},~\forall 0 \leq i \leq r-1\}. %
\end{equation}

The following proposition is a property of the parity-check matrices of binary MDS array codes. %
For more details, please refer to \cite[Ch. 11]{array-code}.

\begin{proposition}
\label{prop:pre-MDS}
Let $\mathcal{C}'$ be the $(k+r, k, l)$ binary array code with the parity-check matrix $\mathbf{H}$ defined in \eqref{eqn:PC-matrix-H}. %
$\mathcal{C}'$ is an MDS code if and only if any $r \times r$ block sub-matrix $\mathbf{H}'$ of the $r \times (k+r)$ block matrix $\mathbf{H}$, denoted as
\begin{equation}
\label{eqn:submatrix-H}
\mathbf{H}' = \begin{bmatrix}
\mathbf{H}_{0,j_0} & \mathbf{H}_{0,j_1} & \cdots & \mathbf{H}_{0, j_{r-1}} \\
\mathbf{H}_{1,j_0} & \mathbf{H}_{1,j_1} & \cdots & \mathbf{H}_{1, j_{r-1}} \\
\vdots & \vdots & \vdots & \vdots \\
\mathbf{H}_{r-1,j_0} & \mathbf{H}_{r-1,j_1} & \cdots & \mathbf{H}_{r-1, j_{r-1}} \\
\end{bmatrix}, 
\end{equation}
where $0 \leq j_0 < j_1 < \cdots < j_{r-1} \leq k+r-1$, is full rank $rl$. 
\end{proposition}

\subsection{The optimal repair/access property}
\label{pre:opt-access}
A $(k+r, k, l)$ binary MDS array code has $k$ information nodes and $r$ parity nodes in each codeword, such that any $k$ out of $k+r$ nodes can reconstruct the whole codeword. %
For single-node failure recovery, the conventional repair approach that downloads all $kl$ bits from any $k$ helper nodes is not efficient in terms of bandwidth since the whole downloaded volume ($kl$ bits)  equals $k$ times the amount of lost data ($l$ bits). %

Assume that a single node becomes unavailable, and the system aims to repair the failed node by connecting to $d$ helper nodes, where $k \leq d \leq k+r-1$. %
By analyzing the information flow graph of storage systems, the seminal work \cite{DSS-repair} demonstrated that for any MDS code, at least $\frac{l}{d-k+1}$ fractions from each of the $d$ helper nodes have to be downloaded to repair a failed node. %
Specific to a $(k+r, k, l)$ binary MDS array code $\mathcal{C}$, for $0 \leq i \leq k+r-1$, let $\mathcal{R}_i$ be a subset of $\{ 0, 1, \cdots, k+r-1 \} \setminus \{i\}$ with cardinality $|\mathcal{R}_i| = d$, and let $\gamma(\mathcal{C}, \mathcal{R}_i)$ be the least number of data bits one needs to download from the helper nodes $\{ \mathbf{c}_j: j \in \mathcal{R}_i \}$ in order to recover the failed node $\mathbf{c}_i$. %
It is well known that
\begin{equation}
\label{eqn:lower-bound}
\gamma(\mathcal{C}, \mathcal{R}_i) \geq \frac{dl}{d-k+1}.
\end{equation}
For each $0 \leq i  \leq k+r-1$, if the lower bound in \eqref{eqn:lower-bound} is achieved when repairing any failed node $\mathbf{c}_i$ using helper nodes $\{ \mathbf{c}_j: j \in \mathcal{R}_i \}$, we say that $\mathcal{C}$ achieves \emph{optimal repair bandwidth} (with respect to a defined $d$). %

During node repair, the downloaded data may be a function of the data stored in these helper nodes. %
Consequently, even codes with optimal repair bandwidth for single-node failures might still require accessing a larger amount of data than the theoretical lower bound in \eqref{eqn:lower-bound}. %
If a failed node of $\mathcal{C}$ can be repaired by accessing an amount of data equal to the theoretical lower bound in \eqref{eqn:lower-bound}, then we say that this node possesses the \emph{optimal access property}. %
Furthermore, for a $(k+r, k, l)$ binary MDS array code $\mathcal{C}$, we say that $\mathcal{C}$ achieves \emph{optimal access bandwidth} (with respect to a defined $d$) if the repair of any single failed node using $d$ helper nodes can be accomplished by accessing only the minimal data volume specified by \eqref{eqn:lower-bound}.  %

\section{Generic construction of binary MDS array codes with optimal access bandwidth}
\label{sec:G-cons-1}

In this section, we propose a generic appraoch for constructing binary MDS array codes with optimal access bandwidth for single-node failures. %
We prove that if the base code is MDS, then the newly constructed code inherently maintains MDS compliance through its algebraic design. %
Furthermore, we rigorously justify the code's optimal access property, demonstrating that for any single-node repair, the repair bandwidth without performing XOR operations at any helper node meets the theoretical lower bound \eqref{eqn:lower-bound},  as reviewed in Sec.\ref{pre:opt-access}.

\subsection{Generic construction}
\label{subsec: G-cons-1}

Consider an arbitrary $(K+r, K, m)$ binary MDS array code defined by its $r \times (K+r)$ parity-check matrix 
\begin{equation}
\label{eqn:Block-PC-matrix-base}
\mathbf{A} = [\mathbf{A}_{i,j}]_{0 \leq i \leq r-1, 0 \leq j \leq K+r-1},
\end{equation}
where every block entry $\mathbf{A}_{i,j}$ of $\mathbf{A}$ is an $m \times m$ matrix over $\mathbb{F}_2$. %
Taking this $(K+r, K, m)$ binary MDS array code as the base code, we next construct a $(k+r, k, l)$ binary MDS array code with optimal access bandwidth. %

\begin{G-construction}
\label{G-construction1}
Given integers $k$ and $s$ satisfying $1 \leq s \leq r$ and $r < k \leq K$, let $l = ms^{\lceil \frac{k+r}{s} \rceil}$ and $g = \lfloor \frac{k+r}{s} \rfloor$. %
Let $\mathcal{C}_1$ denote the $(k+r, k, l)$ binary array code defined by the parity-check matrix $\mathbf{H} = [\mathbf{H}_{i, j}]_{0 \leq i \leq r-1, 0 \leq j \leq k+r-1}$, in which every block entry $\mathbf{H}_{i,j}$ is an $l\times l$ matrix over $\mathbb{F}_2$ constructed based on $\mathbf{A}$ and four arbitrarily chosen $m \times m$ full-rank matrices $\mathbf{\Psi}_1$, $\mathbf{\Psi}_2$, $\mathbf{\Psi}_3$ and $\mathbf{\Psi}_4$ (called coefficient matrices) over $\mathbb{F}_2$ such that the matrix $\left[ \begin{smallmatrix} \mathbf{\Psi}_1 & \mathbf{\Psi}_4 \\ \mathbf{\Psi}_3 & \mathbf{\Psi}_2 \end{smallmatrix} \right]$ is also full rank. %
Regard $\mathbf{H}_{i,j}$ as an $s^{\lceil \frac{k+r}{s} \rceil} \times s^{\lceil \frac{k+r}{s} \rceil}$ block matrix with every block entry an $m\times m$ matrix. %
For $0 \leq a, b \leq s^{\lceil \frac{k+r}{s} \rceil}-1$, the $(a, b)$-th block entry in $\mathbf{H}_{i,j}$, denoted by  $\mathbf{H}_{i, j}(a, b)$, is constructed as follows. Notice that for every $0 \leq j \leq k+r-1$, it can be expressed as $j = vs+u$ for some $0 \leq v \leq g$,  $0 \leq u \leq s-1$. 
\begin{itemize}
  \item For $j = vs+u$ with $0 \leq v \leq g-1$ and $0 \leq u \leq s-1$, 
\begin{equation}
\label{eqn: G-construction2}
\begin{aligned}
\mathbf{H}_{i, j}(a, b) = 
\begin{cases}
\mathbf{A}_{i, j}\mathbf{\Psi}_1 & a_v < u,~b = a \\
\mathbf{A}_{i, j} & a_v = u,~b = a \\
\mathbf{A}_{i, j}\mathbf{\Psi}_2 & a_v > u,~b = a \\
\mathbf{A}_{i, j-u+w}\mathbf{\Psi}_3 & \begin{array}{@{}l@{}} a_v = u,~b = a(v,w),~0 \leq w < u \end{array} \\
\mathbf{A}_{i, j-u+w}\mathbf{\Psi}_4 &   \begin{array}{@{}l@{}} a_v = u,~b = a(v,w),~u < w \leq s-1 \end{array} \\
\mathbf{0} & \text{otherwise} 
\end{cases},
\end{aligned}
\end{equation}
where $a_v$ represents the $v$-th $s$-ary symbol in the $s$-ary expression $\mathbf{s}_a = [a_{g} ~ a_{g-1} ~ \ldots ~ a_0 ]$ of $a$, and $a(v,w)$ represents the integer smaller than $s^{\lceil \frac{k+r}{s} \rceil}$ with $\mathbf{s}_{a(v,w)} = [a_{g} ~ \ldots a_{v+1} ~w ~ a_{v-1}~ \ldots~a_0 ]$.
  \item For $j = gs+u$ with $0 \leq u \leq k+r-sg-1$, 
\begin{equation}
\label{eqn: G-construction2-1}
\begin{aligned}
\mathbf{H}_{i, j}(a, b) = 
\begin{cases}
\mathbf{A}_{i, j}\mathbf{\Psi}_1 & a_v < u,~b = a \\
\mathbf{A}_{i, j} & a_v = u, ~ b = a \\
\mathbf{A}_{i, j}\mathbf{\Psi}_2 & a_v > u,~b = a \\
\mathbf{A}_{i, j-u+w}\mathbf{\Psi}_3 & \begin{array}{@{}l@{}} a_{g} = u,~b = a(g,w),~0 \leq w < u \end{array} \\
\mathbf{A}_{i, \sigma(j-u+w)}\mathbf{\Psi}_4 &   \begin{array}{@{}l@{}} a_{g} = u,~b = a(g,w),~u < w \leq s-1 \end{array} \\
\mathbf{0} & \text{otherwise} 
\end{cases},
\end{aligned}
\end{equation}
where in the case $u < w \leq s-1$, $\sigma(j-u+w)$ denotes $j-u+w \bmod k+r$ for brevity. 
\end{itemize}
\hfill $\blacksquare$
\end{G-construction}

As $g = \lfloor \frac{k+r}{s} \rfloor$, when $s$ divides $k+r$, only \eqref{eqn: G-construction2} is involved in the construction of $\mathcal{C}_1$ constructed by Generic Construction \ref{G-construction1}. %
Moreover, this construction allows flexible selection of coefficient matrices $\mathbf{\Psi}_1, \ldots, \mathbf{\Psi}_4$. A simplified yet valid instantiation can be obtained by setting $\mathbf{\Psi}_1 = \mathbf{\Psi}_2 = \mathbf{\Psi}_3 = \mathbf{I}_m$ and selecting $\mathbf{\Psi}_4$ as any full-rank matrix such that $\mathbf{I}_m + \mathbf{\Psi}_4$ remains full rank. %
To the best of our knowledge, $\mathcal{C}_1$ generally achieves the smallest sub-packetization  $ms^{\lceil \frac{k+r}{s} \rceil}$ among existing binary MDS array codes with optimal access bandwidth, though such sub-packetization is $m$ times the lower bound derived in \cite{lower-bound-22}. %
Notably, as remarked in \cite{Tang-22TCOM-transform}, the lower bound in \cite{lower-bound-22} is not sensitive to binary MDS array codes with optimal access bandwidth. %

\begin{theorem}
\label{theorem: MDS-2}
In Generic Construction \ref{G-construction1}, if the $(K+r, K, m)$ base code is MDS, then the $(k+r, k, ms^{\lceil \frac{k+r}{s} \rceil})$ binary array code $\mathcal{C}_1$ is MDS. %
\end{theorem}

\begin{theorem}
\label{theorem:optimal-access}
In Generic Construction \ref{G-construction1}, if the $(K+r, K, m)$ base code is MDS, then the $(k+r, k, l)$ binary array code $\mathcal{C}_1$ achieves optimal access bandwidth with $d = k+s-1$ helper nodes, among which $s-1$ are designated and $k$ are free to select. %
\end{theorem}

Sec. \ref{subsec:MDS} in the sequel will justify the MDS property of $\mathcal{C}_1$ by proving Theorem \ref{theorem: MDS-2}. %
Subsequently, Sec. \ref{subsec:optaccess-C1} will prove Theorem \ref{theorem:optimal-access} by explicitly specifying the set of the $d = k+s-1$ helper nodes and the exact data bits accessed from each helper node, thereby demonstrating that the repair bandwidth for any single-node failure of $\mathcal{C}_1$ meets the theoretical lower bound \eqref{eqn:lower-bound}. %
To facilitate better understanding of the above construction, we next provide a concrete example to illustrate the structure of the parity-check matrix of code $\mathcal{C}_1$.  

\begin{example}
\label{exam:1}
Set $k = 3$ and $r = s = 2$, so that $s^{\lceil \frac{k+r}{s} \rceil} = 8$. Consider an arbitrary $(5, 3, m)$ binary MDS array code defined by its parity-check matrix $\mathbf{A}$ in the form of \eqref{eqn:Block-PC-matrix-base}. By taking this $(5, 3, m)$ code as a base code, Generic Construction \ref{G-construction1} generates the $(5, 3, 8m)$ code  $\mathcal{C}_1$ with the parity-check matrix $\mathbf{H} = [\mathbf{H}_{i, j}]_{0 \leq i \leq 1, 0 \leq j \leq 4}$ prescribed by %
\begin{equation*}
\small
\begin{aligned}
&\mathbf{H}_{i,0} = 
\begin{bmatrix} 
\mathbf{A}_{i,0} & \mathbf{A}_{i, 1}\mathbf{\Psi}_4 & \mathbf{0} & \mathbf{0} & \mathbf{0} & \mathbf{0} & \mathbf{0} & \mathbf{0} \\
\mathbf{0} & \mathbf{A}_{i,0}\mathbf{\Psi}_2 & \mathbf{0} & \mathbf{0} & \mathbf{0} & \mathbf{0} & \mathbf{0} & \mathbf{0} \\
\mathbf{0} & \mathbf{0} & \mathbf{A}_{i,0} & \mathbf{A}_{i, 1}\mathbf{\Psi}_4 & \mathbf{0} & \mathbf{0} & \mathbf{0} & \mathbf{0} \\
\mathbf{0} & \mathbf{0} & \mathbf{0} & \mathbf{A}_{i,0}\mathbf{\Psi}_2 & \mathbf{0} & \mathbf{0} & \mathbf{0} & \mathbf{0} \\
\mathbf{0} & \mathbf{0} & \mathbf{0} & \mathbf{0} & \mathbf{A}_{i,0} & \mathbf{A}_{i, 1}\mathbf{\Psi}_4 & \mathbf{0} & \mathbf{0} \\
\mathbf{0} & \mathbf{0} & \mathbf{0} & \mathbf{0} & \mathbf{0} & \mathbf{A}_{i,0}\mathbf{\Psi}_2 & \mathbf{0} & \mathbf{0} \\
\mathbf{0} & \mathbf{0} & \mathbf{0} & \mathbf{0} & \mathbf{0} & \mathbf{0} & \mathbf{A}_{i,0} & \mathbf{A}_{i, 1}\mathbf{\Psi}_4 \\
\mathbf{0} & \mathbf{0} & \mathbf{0} & \mathbf{0} & \mathbf{0} & \mathbf{0} & \mathbf{0} & \mathbf{A}_{i,0}\mathbf{\Psi}_2 \\
\end{bmatrix}  \\[3pt]
&\mathbf{H}_{i,1} = 
\begin{bmatrix}
\mathbf{A}_{i, 1}\mathbf{\Psi}_1 & \mathbf{0} & \mathbf{0} & \mathbf{0} & \mathbf{0} & \mathbf{0} & \mathbf{0} & \mathbf{0} \\
\mathbf{A}_{i, 0}\mathbf{\Psi}_3 & \mathbf{A}_{i, 1} & \mathbf{0} & \mathbf{0} & \mathbf{0} & \mathbf{0} & \mathbf{0} & \mathbf{0} \\
\mathbf{0} & \mathbf{0} & \mathbf{A}_{i, 1}\mathbf{\Psi}_1 & \mathbf{0} & \mathbf{0} & \mathbf{0} & \mathbf{0} & \mathbf{0} \\
\mathbf{0} & \mathbf{0} & \mathbf{A}_{i, 0}\mathbf{\Psi}_3 & \mathbf{A}_{i, 1} & \mathbf{0} & \mathbf{0} & \mathbf{0} & \mathbf{0} \\
\mathbf{0} & \mathbf{0} & \mathbf{0} & \mathbf{0} & \mathbf{A}_{i, 1}\mathbf{\Psi}_1 & \mathbf{0} & \mathbf{0} & \mathbf{0} \\
\mathbf{0} & \mathbf{0} & \mathbf{0} & \mathbf{0} & \mathbf{A}_{i, 0}\mathbf{\Psi}_3 & \mathbf{A}_{i, 1} & \mathbf{0} & \mathbf{0} \\
\mathbf{0} & \mathbf{0} & \mathbf{0} & \mathbf{0} & \mathbf{0} & \mathbf{0} & \mathbf{A}_{i, 1}\mathbf{\Psi}_1 & \mathbf{0} \\
\mathbf{0} & \mathbf{0} & \mathbf{0} & \mathbf{0} & \mathbf{0} & \mathbf{0} & \mathbf{A}_{i, 0}\mathbf{\Psi}_3 & \mathbf{A}_{i, 1} \\
\end{bmatrix}\\[3pt]
\end{aligned}
\end{equation*}
\begin{equation*}
\small
\begin{aligned}
&\mathbf{H}_{i,2} = 
\begin{bmatrix}
\mathbf{A}_{i, 2} & \mathbf{0} & \mathbf{A}_{i, 3}\mathbf{\Psi}_4 & \mathbf{0} & \mathbf{0} & \mathbf{0} & \mathbf{0} & \mathbf{0} \\
\mathbf{0} & \mathbf{A}_{i, 2} & \mathbf{0} & \mathbf{A}_{i, 3}\mathbf{\Psi}_4 & \mathbf{0} & \mathbf{0} & \mathbf{0} & \mathbf{0} \\
\mathbf{0} & \mathbf{0} & \mathbf{A}_{i, 2}\mathbf{\Psi}_2 & \mathbf{0} & \mathbf{0} & \mathbf{0} & \mathbf{0} & \mathbf{0} \\
\mathbf{0} & \mathbf{0} & \mathbf{0} & \mathbf{A}_{i, 2}\mathbf{\Psi}_2 & \mathbf{0} & \mathbf{0} & \mathbf{0} & \mathbf{0} \\
\mathbf{0} & \mathbf{0} & \mathbf{0} & \mathbf{0} & \mathbf{A}_{i, 2} & \mathbf{0} & \mathbf{A}_{i, 3}\mathbf{\Psi}_4 & \mathbf{0} \\
\mathbf{0} & \mathbf{0} & \mathbf{0} & \mathbf{0} & \mathbf{0} & \mathbf{A}_{i, 2} & \mathbf{0} & \mathbf{A}_{i, 3}\mathbf{\Psi}_4 \\
\mathbf{0} & \mathbf{0} & \mathbf{0} & \mathbf{0} & \mathbf{0} & \mathbf{0} & \mathbf{A}_{i, 2}\mathbf{\Psi}_2 & \mathbf{0} \\
\mathbf{0} & \mathbf{0} & \mathbf{0} & \mathbf{0} & \mathbf{0} & \mathbf{0} & \mathbf{0} & \mathbf{A}_{i, 2}\mathbf{\Psi}_2 \\
\end{bmatrix}\\[3pt]
&\mathbf{H}_{i,3} = 
\begin{bmatrix}
\mathbf{A}_{i, 3}\mathbf{\Psi}_1 & \mathbf{0} & \mathbf{0} & \mathbf{0} & \mathbf{0} & \mathbf{0} & \mathbf{0} & \mathbf{0} \\
\mathbf{0} & \mathbf{A}_{i, 3}\mathbf{\Psi}_1 & \mathbf{0} & \mathbf{0} & \mathbf{0} & \mathbf{0} & \mathbf{0} & \mathbf{0} \\
\mathbf{A}_{i, 2}\mathbf{\Psi}_3 & \mathbf{0} & \mathbf{A}_{i, 3} & \mathbf{0} & \mathbf{0} & \mathbf{0} & \mathbf{0} & \mathbf{0} \\
\mathbf{0} & \mathbf{A}_{i, 2}\mathbf{\Psi}_3 & \mathbf{0} & \mathbf{A}_{i, 3} & \mathbf{0} & \mathbf{0} & \mathbf{0} & \mathbf{0} \\
\mathbf{0} & \mathbf{0} & \mathbf{0} & \mathbf{0} & \mathbf{A}_{i, 3}\mathbf{\Psi}_1 & \mathbf{0} & \mathbf{0} & \mathbf{0} \\
\mathbf{0} & \mathbf{0} & \mathbf{0} & \mathbf{0} & \mathbf{0} & \mathbf{A}_{i, 3}\mathbf{\Psi}_1 & \mathbf{0} & \mathbf{0} \\
\mathbf{0} & \mathbf{0} & \mathbf{0} & \mathbf{0} & \mathbf{A}_{i, 2}\mathbf{\Psi}_3 & \mathbf{0} & \mathbf{A}_{i, 3} & \mathbf{0} \\
\mathbf{0} & \mathbf{0} & \mathbf{0} & \mathbf{0} & \mathbf{0} & \mathbf{A}_{i, 2}\mathbf{\Psi}_3 & \mathbf{0} & \mathbf{A}_{i, 3} \\
\end{bmatrix} \\[3pt]
&\mathbf{H}_{i,4} = 
\begin{bmatrix}
\mathbf{A}_{i, 4} & \mathbf{0} & \mathbf{0} & \mathbf{0} & \mathbf{A}_{i, 0}\mathbf{\Psi}_4 & \mathbf{0} & \mathbf{0} & \mathbf{0} \\
\mathbf{0} & \mathbf{A}_{i, 4} & \mathbf{0} & \mathbf{0} & \mathbf{0} & \mathbf{A}_{i, 0}\mathbf{\Psi}_4  & \mathbf{0} & \mathbf{0} \\
\mathbf{0} & \mathbf{0} & \mathbf{A}_{i, 4} & \mathbf{0} & \mathbf{0} & \mathbf{0} & \mathbf{A}_{i, 0}\mathbf{\Psi}_4  & \mathbf{0} \\
\mathbf{0} & \mathbf{0} & \mathbf{0} & \mathbf{A}_{i, 4} & \mathbf{0} & \mathbf{0} & \mathbf{0} & \mathbf{A}_{i, 0}\mathbf{\Psi}_4  \\
\mathbf{0} & \mathbf{0} & \mathbf{0} & \mathbf{0} & \mathbf{A}_{i, 4}\mathbf{\Psi}_2 & \mathbf{0} & \mathbf{0} & \mathbf{0} \\
\mathbf{0} & \mathbf{0} & \mathbf{0} & \mathbf{0} & \mathbf{0} & \mathbf{A}_{i, 4}\mathbf{\Psi}_2 & \mathbf{0} & \mathbf{0} \\
\mathbf{0} & \mathbf{0} & \mathbf{0} & \mathbf{0} & \mathbf{0} & \mathbf{0} & \mathbf{A}_{i, 4}\mathbf{\Psi}_2 & \mathbf{0} \\
\mathbf{0} & \mathbf{0} & \mathbf{0} & \mathbf{0} & \mathbf{0} & \mathbf{0} & \mathbf{0} & \mathbf{A}_{i, 4}\mathbf{\Psi}_2 \\
\end{bmatrix} \\
\end{aligned}.
\end{equation*}%
\hfill $\blacksquare$
\end{example}

Notice that the codes in \cite{Hou-20TIT-transform} and \cite{Tang-22TCOM-transform} are also constructed based on a base code and share a similar  sub-packetization level with our code $\mathcal{C}_1$. %
Taking an arbitrary $(k+r, k, m)$ binary MDS array code with even sub-packetization $m$ as the base code, the construction in \cite{Tang-22TCOM-transform} yields a $(k+r, k, ms^{\lceil \frac{k+r}{s} \rceil})$ code with optimal access bandwidth for the case $s = r$. %
In addition, for any given $s \leq r$, the construction in \cite{Hou-20TIT-transform} utilizes the $(k+r, k, m)$ EVENODD code \cite{Blaum-Evenodd-TComP95} \cite{Blaum-TIT96-MDS-Generalized_EvenOdd} (where $m+1$ is a prime number and $k \leq m+1$) as the base code to generate a $(k+r, k, ms^{\lceil \frac{k}{s} \rceil + \lceil \frac{r}{s} \rceil})$ code with optimal access bandwidth. %
Through the following concrete example, we demonstrate the architectural differences manifested in the parity-check matrices between code $\mathcal{C}_1$ and the codes in \cite{Hou-20TIT-transform} and \cite{Tang-22TCOM-transform}. %

\begin{example}
\label{exam:2}
Set $k = 3$ and $r = s = 2$, so that $s^{\lceil \frac{k+r}{s} \rceil} = s^{\lceil \frac{k}{s} \rceil + \lceil \frac{r}{s} \rceil} = 8$. Consider an arbitrary $(5, 3, m)$ binary MDS array code defined by its parity-check matrix $\mathbf{A}$ in the form of \eqref{eqn:Block-PC-matrix-base}. %
By taking this $(5, 3, m)$ code as a base code, a $(5, 3, 8m)$ binary array code can be generated with the parity-check matrix $\mathbf{H} = [\mathbf{H}_{i, j}]_{0 \leq i \leq 1, 0 \leq j \leq 4}$ prescribed by   
\begin{equation*}
\small
\begin{aligned}
&\mathbf{H}_{i,0} = 
\begin{bmatrix} 
\mathbf{A}_{i,0} & \mathbf{A}_{i, 1}\mathbf{\Lambda} & \mathbf{0} & \mathbf{A}_{i, 2}\mathbf{\Lambda} & \mathbf{0} & \mathbf{0} & \mathbf{0} & \mathbf{0} \\
\mathbf{0} & \mathbf{A}_{i,0}\mathbf{\Lambda} & \mathbf{0} & \mathbf{0} & \mathbf{0} & \mathbf{0} & \mathbf{0} & \mathbf{0} \\
\mathbf{0} & \mathbf{0} & \mathbf{A}_{i,0} & \mathbf{A}_{i, 1}\mathbf{\Lambda} & \mathbf{0} & \mathbf{0} & \mathbf{0} & \mathbf{0} \\
\mathbf{0} & \mathbf{0} & \mathbf{0} & \mathbf{A}_{i,0}\mathbf{\Lambda} & \mathbf{0} & \mathbf{0} & \mathbf{0} & \mathbf{0} \\
\mathbf{0} & \mathbf{0} & \mathbf{0} & \mathbf{0} & \mathbf{A}_{i,0} & \mathbf{A}_{i, 1}\mathbf{\Lambda} & \mathbf{0} & \mathbf{A}_{i, 2}\mathbf{\Lambda} \\
\mathbf{0} & \mathbf{0} & \mathbf{0} & \mathbf{0} & \mathbf{0} & \mathbf{A}_{i,0}\mathbf{\Lambda} & \mathbf{0} & \mathbf{0} \\
\mathbf{0} & \mathbf{0} & \mathbf{0} & \mathbf{0} & \mathbf{0} & \mathbf{0} & \mathbf{A}_{i,0} & \mathbf{A}_{i, 1}\mathbf{\Lambda} \\
\mathbf{0} & \mathbf{0} & \mathbf{0} & \mathbf{0} & \mathbf{0} & \mathbf{0} & \mathbf{0} & \mathbf{A}_{i,0}\mathbf{\Lambda} \\
\end{bmatrix}  \\
&\mathbf{H}_{i,1} = 
\begin{bmatrix}
\mathbf{A}_{i, 1}\mathbf{\Lambda} & \mathbf{0} & \mathbf{A}_{i, 2}\mathbf{\Lambda} & \mathbf{0} & \mathbf{0} & \mathbf{0} & \mathbf{0} & \mathbf{0} \\
\mathbf{A}_{i, 0}\bar{\mathbf{\Lambda}} & \mathbf{A}_{i, 1} & \mathbf{0} & \mathbf{A}_{i, 2}\mathbf{\Lambda} & \mathbf{0} & \mathbf{0} & \mathbf{0} & \mathbf{0} \\
\mathbf{0} & \mathbf{0} & \mathbf{A}_{i, 1}\mathbf{\Lambda} & \mathbf{0} & \mathbf{0} & \mathbf{0} & \mathbf{0} & \mathbf{0} \\
\mathbf{0} & \mathbf{0} & \mathbf{A}_{i, 0}\bar{\mathbf{\Lambda}} & \mathbf{A}_{i, 1} & \mathbf{0} & \mathbf{0} & \mathbf{0} & \mathbf{0} \\
\mathbf{0} & \mathbf{0} & \mathbf{0} & \mathbf{0} & \mathbf{A}_{i, 1}\mathbf{\Lambda}& \mathbf{0} & \mathbf{A}_{i, 2}\mathbf{\Lambda} & \mathbf{0} \\
\mathbf{0} & \mathbf{0} & \mathbf{0} & \mathbf{0} & \mathbf{A}_{i, 0}\bar{\mathbf{\Lambda}} & \mathbf{A}_{i, 1} & \mathbf{0} & \mathbf{A}_{i, 2}\mathbf{\Lambda} \\
\mathbf{0} & \mathbf{0} & \mathbf{0} & \mathbf{0} & \mathbf{0} & \mathbf{0} & \mathbf{A}_{i, 1}\mathbf{\Lambda} & \mathbf{0} \\
\mathbf{0} & \mathbf{0} & \mathbf{0} & \mathbf{0} & \mathbf{0} & \mathbf{0} & \mathbf{A}_{i, 0}\bar{\mathbf{\Lambda}} & \mathbf{A}_{i, 1} \\
\end{bmatrix}\\
&\mathbf{H}_{i,2} = 
\begin{bmatrix}
\mathbf{A}_{i, 2}\mathbf{\Lambda} & \mathbf{0} & \mathbf{0} & \mathbf{0} & \mathbf{0} & \mathbf{0} & \mathbf{0} & \mathbf{0} \\
\mathbf{0} & \mathbf{A}_{i, 2}\mathbf{\Lambda} & \mathbf{0} & \mathbf{0} & \mathbf{0} & \mathbf{0} & \mathbf{0} & \mathbf{0} \\
\mathbf{A}_{i, 1}\bar{\mathbf{\Lambda}} & \mathbf{0} & \mathbf{A}_{i, 2} & \mathbf{0} & \mathbf{0} & \mathbf{0} & \mathbf{0} & \mathbf{0} \\
\mathbf{0} & \mathbf{A}_{i, 1}\bar{\mathbf{\Lambda}} & \mathbf{0} & \mathbf{A}_{i, 2} & \mathbf{0} & \mathbf{0} & \mathbf{0} & \mathbf{0} \\
\mathbf{0} & \mathbf{0} & \mathbf{0} & \mathbf{0} & \mathbf{A}_{i, 2}\mathbf{\Lambda} & \mathbf{0} & \mathbf{0} & \mathbf{0} \\
\mathbf{0} & \mathbf{0} & \mathbf{0} & \mathbf{0} & \mathbf{0} & \mathbf{A}_{i, 2}\mathbf{\Lambda} & \mathbf{0} & \mathbf{0} \\
\mathbf{0} & \mathbf{0} & \mathbf{0} & \mathbf{0} & \mathbf{A}_{i, 1}\bar{\mathbf{\Lambda}} & \mathbf{0} & \mathbf{A}_{i, 2} & \mathbf{0} \\
\mathbf{0} & \mathbf{0} & \mathbf{0} & \mathbf{0} & \mathbf{0} & \mathbf{A}_{i, 1}\bar{\mathbf{\Lambda}} & \mathbf{0} & \mathbf{A}_{i, 2} \\
\end{bmatrix}\\
\end{aligned}
\end{equation*}
\begin{equation*}
\small
\begin{aligned}
&\mathbf{H}_{i,3} = 
\begin{bmatrix}
\mathbf{A}_{i, 3} & \mathbf{0} & \mathbf{0} & \mathbf{0} & \mathbf{A}_{i,4}\mathbf{\Lambda} & \mathbf{0} & \mathbf{0} & \mathbf{0} \\
\mathbf{0} & \mathbf{A}_{i, 3} & \mathbf{0} & \mathbf{0} & \mathbf{0} & \mathbf{A}_{i,4}\mathbf{\Lambda} & \mathbf{0} & \mathbf{0} \\
\mathbf{0} & \mathbf{0} & \mathbf{A}_{i, 3} & \mathbf{0} & \mathbf{0} & \mathbf{0} & \mathbf{A}_{i,4}\mathbf{\Lambda} & \mathbf{0} \\
\mathbf{0} & \mathbf{0} & \mathbf{0} & \mathbf{A}_{i, 3} & \mathbf{0} & \mathbf{0} & \mathbf{0} & \mathbf{A}_{i,4}\mathbf{\Lambda} \\
\mathbf{0} & \mathbf{0} & \mathbf{0} & \mathbf{0} & \mathbf{A}_{i, 3}\mathbf{\Lambda} & \mathbf{0} & \mathbf{0} & \mathbf{0} \\
\mathbf{0} & \mathbf{0} & \mathbf{0} & \mathbf{0} & \mathbf{0} & \mathbf{A}_{i, 3}\mathbf{\Lambda} & \mathbf{0} & \mathbf{0} \\
\mathbf{0} & \mathbf{0} & \mathbf{0} & \mathbf{0} & \mathbf{0} & \mathbf{0} & \mathbf{A}_{i, 3}\mathbf{\Lambda} & \mathbf{0} \\
\mathbf{0} & \mathbf{0} & \mathbf{0} & \mathbf{0} & \mathbf{0} & \mathbf{0} & \mathbf{0} & \mathbf{A}_{i, 3}\mathbf{\Lambda} \\
\end{bmatrix} \\
&\mathbf{H}_{i,4} = 
\begin{bmatrix}
\mathbf{A}_{i, 4}\mathbf{\Lambda} & \mathbf{0} & \mathbf{0} & \mathbf{0} & \mathbf{0} & \mathbf{0} & \mathbf{0} & \mathbf{0} \\
\mathbf{0} & \mathbf{A}_{i, 4}\mathbf{\Lambda} & \mathbf{0} & \mathbf{0} & \mathbf{0} & \mathbf{0}  & \mathbf{0} & \mathbf{0} \\
\mathbf{0} & \mathbf{0} & \mathbf{A}_{i, 4}\mathbf{\Lambda} & \mathbf{0} & \mathbf{0} & \mathbf{0} & \mathbf{0}  & \mathbf{0} \\
\mathbf{0} & \mathbf{0} & \mathbf{0} & \mathbf{A}_{i, 4}\mathbf{\Lambda} & \mathbf{0} & \mathbf{0} & \mathbf{0} & \mathbf{0}  \\
\mathbf{A}_{i, 3}\bar{\mathbf{\Lambda}} & \mathbf{0} & \mathbf{0} & \mathbf{0} & \mathbf{A}_{i, 4} & \mathbf{0} & \mathbf{0} & \mathbf{0} \\
\mathbf{0} & \mathbf{A}_{i, 3}\bar{\mathbf{\Lambda}} & \mathbf{0} & \mathbf{0} & \mathbf{0} & \mathbf{A}_{i, 4} & \mathbf{0} & \mathbf{0} \\
\mathbf{0} & \mathbf{0} & \mathbf{A}_{i, 3}\bar{\mathbf{\Lambda}} & \mathbf{0} & \mathbf{0} & \mathbf{0} & \mathbf{A}_{i, 4} & \mathbf{0} \\
\mathbf{0} & \mathbf{0} & \mathbf{0} & \mathbf{A}_{i, 3}\bar{\mathbf{\Lambda}} & \mathbf{0} & \mathbf{0} & \mathbf{0} & \mathbf{A}_{i, 4} \\
\end{bmatrix} \\
\end{aligned},
\end{equation*}%
where the coefficient matrices $\mathbf{\Lambda}$ and $\bar{\mathbf{\Lambda}}$ are certain $m\times m$ binary matrices. %
Under the setting of even $m$, $\mathbf{\Lambda} = \left[ \begin{smallmatrix} \mathbf{I}_{m/2} & \mathbf{I}_{m/2} \\ \mathbf{I}_{m/2} & \mathbf{0} \end{smallmatrix} \right]$, and $\bar{\mathbf{\Lambda}} = \mathbf{I}_m + \mathbf{\Lambda}$, this $(5, 3, 8m)$ code coincides with the MDS code with optimal access bandwidth constructed in 
\cite{Tang-22TCOM-transform}. %
Next assume the $(5, 3, m)$ base code is an EVENODD code defined by the parity-check matrix 
\begin{equation*}
\mathbf{A} = [ \mathbf{A}_{i, j}]_{0 \leq i \leq 1, 0 \leq j \leq 4}  = 
\begin{bmatrix} 
\mathbf{I}_m & \mathbf{I}_m & \mathbf{I}_m & \mathbf{I}_m & \mathbf{0} \\ 
\mathbf{I}_m & \mathbf{P}\mathbf{C}_{m+1}^{m}\mathbf{Q} & \mathbf{P}\mathbf{C}_{m+1}^{m-1}\mathbf{Q} & \mathbf{0} & \mathbf{I}_m
\end{bmatrix},  
\end{equation*}
where $\mathbf{P} = [\mathbf{I}_m~\mathbf{1}]$ is an $m \times (m+1)$ matrix, $\mathbf{C}_{m+1} = \left[ \begin{smallmatrix} \mathbf{0} ~ \mathbf{I}_m \\ 1 ~ \mathbf{0} \end{smallmatrix} \right]$ is an $(m+1) \times (m+1)$ circulant matrix  and $\mathbf{Q} = [\mathbf{I}_m ~ \mathbf{0}]^{\rm T}$ is an $(m+1) \times m$ matrix. Under the setting of $\mathbf{\Lambda} = \mathbf{P}\mathbf{C}_{m+1}\mathbf{Q}$ and $\bar{\mathbf{\Lambda}} = \mathbf{P}(\mathbf{I}_{m+1} + \mathbf{C}_{m+1})\mathbf{Q}$, the constructed $(5, 3, 8m)$ code coincides with the MDS code with optimal access bandwidth constructed in \cite{Hou-20TIT-transform}.
\hfill $\blacksquare$
\end{example}

One can observe that, for the same $(5, 3, m)$ base code, regardless of the choice of its parity-check matrix, the $(5, 3, 8m)$ codes illustrated in Example \ref{exam:2} share an identical parity-check matrix structure, differing only in the settings of $\mathbf{\Lambda}$ and $\bar{\mathbf{\Lambda}}$. %
In contrast, the $(5, 3, 8m)$ code in Example \ref{exam:1} has a distinct parity-check matrix, since some positions of the nonzero block matrices differ from those in Example \ref{exam:2}, and there is more flexibility for block matrix selection (that is, the flexible setting of coefficient matrices $\mathbf{\Psi}_1, \ldots, \mathbf{\Psi}_4$). %
We thus demonstrated that in the special case $s = r$ and $s^{\lceil \frac{k+r}{s} \rceil} = s^{\lceil \frac{k}{s} \rceil + \lceil \frac{r}{s} \rceil} $, even though the new code $\mathcal{C}_1$ constructed by Generic Construction \ref{G-construction1} has the same sub-packetization as those constructed in  \cite{Hou-20TIT-transform} and \cite{Tang-22TCOM-transform}, it has a different structure. %
In addition, the construction in \cite{Tang-22TCOM-transform} requires the sub-packetization $m$ of the base code to be even, and the construction in \cite{Hou-20TIT-transform} requires that $m+1$ be prime with $m+1 \geq k$, while Generic Construction \ref{G-construction1} has no such constraint. %
If the sub-packetization $m$ of the base code is odd in \cite{Tang-22TCOM-transform}, two base code instances are combined in advance to ensure even sub-packetization, which doubles the sub-packetization of the constructed code with optimal access bandwidth from $mr^{\lceil \frac{k+r}{r} \rceil}$ to $2mr^{\lceil \frac{k+r}{r} \rceil}$. %

It is also worthwhile to notice that Generic Construction \ref{G-construction1} also applies to the general case $s \leq r$.  
In comparison, the construction in \cite{Hou-20TIT-transform} also covers this regime but is limited to selecting an EVENODD code as the base code. %
The construction in \cite{Tang-22TCOM-transform}, similar to Generic Construction \ref{G-construction1}, utilizes an arbitrary binary MDS array code as the base code. However, such construction was only investigated in \cite{Tang-22TCOM-transform} for the case $s = r$. %
Although \cite{Tang-22TCOM-transform} claimed that the construction can be extended to $s \leq r$ in a straightforward way, the resulting sub-packetization for  the general case is $ms^{\lceil \frac{k}{s} \rceil+\lceil \frac{r}{s}\rceil}$, which is generally larger than the sub-packetization $ms^{\lceil \frac{k+r}{s} \rceil}$ of $\mathcal{C}_1$ proposed in this section. %

\subsection{Justification of the MDS property of $\mathcal{C}_1$}
\label{subsec:MDS}
This subsection will justify the MDS property of the $(k+r, k, ms^{\lceil \frac{k+r}{s} \rceil})$ code $\mathcal{C}_1$ constructed by Generic Construction \ref{G-construction1}, so that Theorem \ref{theorem: MDS-2} can be proved.

We first select a $(K+r, K, m)$ binary MDS array code as the base code. %
Recall that for defined $k$ and $s$, $l = ms^{\lceil \frac{k+r}{s} \rceil}$. Let $l' = s^{\lceil \frac{k+r}{s} \rceil}$. %
Consider arbitrary $j_0, j_1, \ldots, j_{r-1}$ subject to  $0 \leq j_0 < j_1 < \ldots < j_{r-1} \leq k+r-1$, and let $\mathbf{H}'$ denote the $r \times r$ block sub-matrix of $\mathbf{H}$ as defined in \eqref{eqn:submatrix-H}, that is, $\mathbf{H}'$ is obtained from $\mathbf{H}$ by restricting to the block columns indexed by $j_0, j_1, \ldots, j_{r-1}$. %
In order to prove the MDS property of $\mathcal{C}_1$, according to Proposition \ref{prop:pre-MDS}, it is equivalent to prove that $\mathbf{H}'$, when regarded as an $rl \times rl$ matrix over $\mathbb{F}_2$, is full rank. %
To establish full rank of $\mathbf{H}'$, we shall show that for an $rl$-bit column vector $\mathbf{x}$, 
\begin{equation}
\label{eqn:H'x = 0}
\mathbf{H}'\mathbf{x} = \mathbf{0}~\mathrm{implies}~\mathbf{x} = \mathbf{0}.
\end{equation}
Here, $\mathbf{x}$ is expressed as $\mathbf{x} = [\mathbf{x}_0^{\rm T}~\mathbf{x}_1^{\rm T}~\ldots~\mathbf{x}_{rl'-1}^{\rm T}]^{\rm T}$, where each $\mathbf{x}_i$ is an $m$-bit column vector. %

Define a block permutation matrix $\mathbf{P} = [\mathbf{P}_{i,j}]_{0 \leq i, j \leq rl'-1}$ such that
\begin{equation}
\label{eqn:P-MDS}
\mathbf{P}_{i,j} = \mathbf{I}_m ~{\rm iff}~i = rj - (rl'-1)\left\lfloor \frac{j}{l'} \right\rfloor.
\end{equation}
Let $\mathbf{H}'' = \mathbf{P}\mathbf{H}'$. In order to prove \eqref{eqn:H'x = 0}, it is equivalent to show that 
\begin{equation}
\label{eqn:mK''=0}
\mathbf{H}''\mathbf{x} = \mathbf{0}~\mathrm{implies}~\mathbf{x} = \mathbf{0}.
\end{equation}
Since each $\mathbf{x}_i$ in $\mathbf{x}$ is an $m$-bit vector, we regard $\mathbf{H}''$ as an $rl' \times rl'$ block matrix, where each block entry is an $m \times m$ matrix over $\mathbb{F}_2$. %
For each $0 \leq a \leq l'-1 $, define $\mathbf{H}''^{(a)}$ as the $r \times rl'$ block sub-matrix of $\mathbf{H}''$ consisting of the block rows indexed by $ar, ar+1, \ldots, ar+r-1$. %
For each $0 \leq j \leq (g+1)s$, we further define a block column vector 
\begin{equation}
\label{eqn:L}
\mathbf{L}_{j} = \begin{bmatrix} \mathbf{A}_{0,\sigma(j)}^{\rm T} ~ \mathbf{A}_{1,\sigma(j)}^{\rm T}~ \ldots ~ \mathbf{A}_{r-1,\sigma(j)}^{\rm T} \end{bmatrix}^{\rm T}. %
\end{equation}
Recall that $\sigma(j)$ denotes $j \bmod k+r$. %
Consequently, from \eqref{eqn: G-construction2} and \eqref{eqn: G-construction2-1}, for each $0 \leq a \leq l'-1$, all the nonzero block columns of $\mathbf{H}''^{(a)}$ belong to the set
\begin{equation*}
\begin{aligned}
&\{ 
\mathbf{L}_{0}\mathbf{\Psi}, \ldots, \mathbf{L}_{k+r-1}\mathbf{\Psi}, \mathbf{L}_{k+r}\mathbf{\Psi}, \ldots, \mathbf{L}_{(g+1)s-1}\mathbf{\Psi}~: \mathbf{\Psi} \in \{ \mathbf{I}_m, \mathbf{\Psi}_1, \mathbf{\Psi}_2, \mathbf{\Psi}_3, \mathbf{\Psi}_4 \}
\}.
\end{aligned}
\end{equation*}
Notice that only when $s$ does not divide $k+r$ and the set $\{ gs, gs+1, \ldots, k+r-1 \}$ has non-empty intersection with $\{j_0, j_1, \ldots, j_{r-1}\}$, some of the nonzero block columns of $\mathbf{H}''^{(a)}$ belong to the set $\{ \mathbf{L}_{k+r}\mathbf{\Psi}, \mathbf{L}_{k+r+1}\mathbf{\Psi}, \ldots, \mathbf{L}_{(g+1)s-1}\mathbf{\Psi}~:~\mathbf{\Psi} \in \{ \mathbf{I}_m, \mathbf{\Psi}_1, \mathbf{\Psi}_2, \mathbf{\Psi}_3, \mathbf{\Psi}_4 \} \}$. %

We now define subsets of block column indices for block matrix $\mathbf{H}''^{(a)}$ for each $0 \leq a \leq l'-1$ :
\begin{itemize}
  \item $\mathcal{U}^{(a)} = \{0 \leq j \leq (g+1)s-1: \mathbf{H}''^{(a)}~\mathrm{contains~a~nonzero~block~column~belonging~to~} \{\mathbf{L}_j, \mathbf{L}_j\mathbf{\Psi}_1, \mathbf{L}_j\mathbf{\Psi}_2, \mathbf{L}_j\mathbf{\Psi}_3, \\ \mathbf{L}_j\mathbf{\Psi}_4 \}\}$; 
  \item $\mathcal{J}^{(a)} \subset \{0, 1, \ldots, rl'-1\}$ lists the indices of nonzero block columns in $\mathbf{H}''^{(a)}$;
  \item For each $0 \leq j \leq (g+1)s-1$, $\mathcal{J}^{(a)}(j) = \{t \in \mathcal{J}^{(a)}: t\textrm{-}\mathrm{th}~\mathrm{block~column~of}~\mathbf{H}''^{(a)}~ \mathrm{belongs~to} \{\mathbf{L}_j, \mathbf{L}_j\mathbf{\Psi}_1, \\ \mathbf{L}_j\mathbf{\Psi}_2, \mathbf{L}_j\mathbf{\Psi}_3, \mathbf{L}_j\mathbf{\Psi}_4 \} \}$. %
\end{itemize}

It is clear that even though the sets $\mathcal{J}^{(a)}$, $0 \leq a \leq l'-1$, are not disjoint, their union is nonetheless the complete set: 
\begin{equation}
\label{eqn:J_union}
\bigcup_{0 \leq a \leq l'-1} \mathcal{J}^{(a)} = \{0, 1, \ldots, rl'-1\}.
\end{equation}
Moreover, the sets $\mathcal{J}^{(a)}(j)$ (each containing one or two indices) form a partition of the set $\mathcal{J}^{(a)}$, so
\begin{equation}
\label{eqn:J_a_partition}
\mathcal{J}^{(a)} = \bigcup_{j \in \mathcal{U}^{(a)}} \mathcal{J}^{(a)}(j) . 
\end{equation}

The next lemma justifies that there is at least one set $\mathcal{U}^{(a)}$ that contains exactly $r$ elements.
\begin{lemma}
\label{lemma:U=r}
For any $0 \leq j_0 < j_1 < \ldots < j_{r-1} \leq k+r-1$, 
\begin{equation}
\label{eqn:appen-proof-MDS2-00}
\min\limits_{a \in  \{0, 1, \ldots, l'-1 \} } |\mathcal{U}^{(a)}| = r.
\end{equation}
\begin{IEEEproof}
Please refer to Appendix-\ref{appendix:lemma-U=r}. %
\end{IEEEproof}
\end{lemma}

Furthermore, Lemma \ref{lemma:U=r} allows us to prove the next lemma by induction on the cardinality of the set $\mathcal{U}^{(a)}$. %

\begin{lemma}
\label{lemma:J_a}
For each $0 \leq a \leq l'-1$, $\mathbf{H}''^{(a)}[\mathbf{x}_0^{\rm T}~\mathbf{x}_1^{\rm T}~\ldots~\mathbf{x}_{rl'-1}^{\rm T}]^{\rm T} = \mathbf{0}$ implies that $\mathbf{x}_t = \mathbf{0}$ for all $t \in \mathcal{J}^{(a)}$. %
\begin{IEEEproof}
Please refer to Appendix-\ref{appendix:lemma-J_a}. %
\end{IEEEproof}
\end{lemma} 

Since the row-wise juxtaposition of $\mathbf{H}''^{(a)}$, $0 \leq a \leq l'-1$, forms $\mathbf{H}''$, and $\bigcup_{0 \leq a \leq l'-1} \mathcal{J}^{(a)} = \{0, 1, \ldots, rl'-1\}$, %
Lemma \ref{lemma:J_a} further implies \eqref{eqn:mK''=0}. We have thus proved that if the base code is MDS, then the code $\mathcal{C}_1$ is MDS too. %

\subsection{Justification of the optimal access property of $\mathcal{C}_1$}
\label{subsec:optaccess-C1}

Consider the $(k+r, k, l = ms^{\lceil \frac{k+r}{s} \rceil})$ binary MDS array code $\mathcal{C}_1$ constructed by Generic Construction \ref{G-construction1}. %
This subsection will justify the optimal access property of $\mathcal{C}_1$, so that Theorem \ref{theorem:optimal-access} can be proved.
Code $\mathcal{C}_1$ is said to have the optimal access property if any failed node can be repaired by connecting to $d = k+s-1$ helper nodes and by accessing exactly $\frac{l}{s}$ bits from each helper node. %
In the following, we prove that any failed node of $\mathcal{C}_1$ can be recovered by accessing exactly $\frac{l}{s}$ bits from each of the $d$ helper nodes, in which $s-1$ are designated and $k$ are free to select. %
This, in turn, implies that code $\mathcal{C}_1$ possesses the optimal access property. %
Throughout this section, unless otherwise specified, $u$ is an arbitrary integer between $0$ and $s-1$, and $v$ is an arbitrary integer between $0$ and $g$. %
We shall specify the amount of data bits accessed from all helper nodes during the repair of  node $\mathbf{c}_j$ with $j = vs+u$. 

Let $l' = s^{\lceil \frac{k+r}{s} \rceil}$. %
For each $0 \leq t \leq k+r-1$, represent the $t$-th node $\mathbf{c}_t$ as $\mathbf{c}_t = [\mathbf{c}_{t,0}^{\rm T} ~ \mathbf{c}_{t,1}^{\rm T} ~\cdots~ \mathbf{c}_{t,l'-1}^{\rm T}]^{\rm T}$, where each data chunk $\mathbf{c}_{t, i}$ is an $m$-bit column vector, and define the following data chunk set of the $t$-th node:
\begin{equation}
\label{eqn:M_coordinate}
\mathcal{M}_t^{(v, u)} =  \big\{ \mathbf{c}_{t, a}:~ 0 \leq a \leq l'-1, ~a_v = u \big\},
\end{equation}
where $a_v$ represents the $v$-th $s$-ary symbol in the $s$-ary expression $\mathbf{s}_a = [a_{g} ~ a_{g-1} ~ \cdots ~ a_0 ]$ of $a$. %

Let $\mathcal{N}^{(v)} = \{ \sigma(vs+w): w \in \{ 0, 1, \cdots, s-1\}\}$ and $\bar{\mathcal{N}}^{(v)} = \{ 0, 1, \cdots, k+r-1 \} \setminus \mathcal{N}^{(v)}$, where $\sigma(vs+w)$ denotes $vs+w \bmod k+r$. %
Further, corresponding to every subset $\mathcal{G}^{(v)} \subseteq \bar{\mathcal{N}}^{(v)}$ with cardinality $|\mathcal{G}^{(v)}| = k$, define $\bar{\mathcal{G}}^{(v,u)}$ with cardinality $d$ as 
\begin{equation}
\label{eqn: def_idx_G}
\bar{\mathcal{G}}^{(v,u)} = \mathcal{G}^{(v)} \cup (\mathcal{N}^{(v)} \setminus \{vs+u\}). 
\end{equation}

In order to repair any failed node $\mathbf{c}_j$ with $j = vs+u$ of $\mathcal{C}_1$, it is necessary to connect to $d = k+s-1$ helper nodes and access exactly $\frac{l}{s}$ bits from each helper node. %
Specifically, all $d$ helper nodes are from the set $\{ \mathbf{c}_i~:~i \in \bar{\mathcal{G}}^{(v, u)}\}$, with $s-1$ nodes belonging to $\{\mathbf{c}_i~:~i \in \mathcal{N}^{(v)} \setminus \{vs+u\} \}$ and the other $k$ nodes chosen arbitrarily from $\{\mathbf{c}_i~:~i \in \bar{\mathcal{N}}^{(v)}\}$. %
Furthermore, during the repair of node $\mathbf{c}_j$, the $\frac{l'}{s}$ data chunks accessed from any helper node $\mathbf{c}_i$, $i \in \bar{\mathcal{G}}^{(v, u)}$, correspond exactly to the set $\mathcal{M}_i^{(v, u)}$ defined in \eqref{eqn:M_coordinate}, and these data chunks collectively contain exactly $\frac{l}{s}$ bits. %
Consequently, the following theorem specifies the total amount of data accessed from all helper nodes during the repair of node $\mathbf{c}_j$. %

\begin{theorem}
\label{theorem:Cons2-access-repair}
For the code $\mathcal{C}_1$ constructed by Generic Construction \ref{G-construction1}, node $\mathbf{c}_{j}$ with $j = vs+u$, $0 \leq v \leq g$ and $0 \leq u \leq s-1$, can be recovered by accessing the $\frac{dl'}{s}$ data chunks in the set 
\begin{equation}
\label{node:cons1}
\mathcal{M}^{(v, u)} = \bigcup\nolimits_{i \in \bar{\mathcal{G}}^{(v,u)} } \mathcal{M}_i^{(v, u)}. %
\end{equation}
\begin{IEEEproof}
Please refer to Appendix-\ref{appendix:proof-Cons2-access-repair}. 
\end{IEEEproof}
\end{theorem}

The optimal access property of $\mathcal{C}_1$ refers to the fact that for any single-node failure, the repair process requires connecting to $d = k+s-1$ helper nodes and accessing exactly $\frac{l}{s}$ bits from each helper node, as justified by Theorem \ref{theorem:Cons2-access-repair}.

\begin{example}
Consider the $(5,3, 8m)$ code $\mathcal{C}_1$ presented in Example \ref{exam:1}. %
We have $s = 2$ and thus $d = k+s-1 = 4$. %
Any failed node can be repaired by accessing all $4$ remaining helper nodes. %
Suppose that node $\mathbf{c}_2$ of $\mathcal{C}_1$ fails. %
One can easily observe that the data chunks $\{ \mathbf{c}_{2,0}, \mathbf{c}_{2, 2}\}$ of $\mathbf{c}_2$ correspond to the first block rows of the two parity-check equations defined in \eqref{eqn:def-C}, which can be expressed as
\begin{equation*}
\begin{bmatrix}
\mathbf{A}_{0,2} & \mathbf{A}_{0,3}\mathbf{\Psi}_4 \\
\mathbf{A}_{1,2} & \mathbf{A}_{1,3}\mathbf{\Psi}_4 
\end{bmatrix}
\begin{bmatrix}
\mathbf{c}_{2, 0} \\ \mathbf{c}_{2, 2}
\end{bmatrix} = \begin{bmatrix} \mathbf{C}_0 \\ \mathbf{C}_1 \end{bmatrix}, 
\end{equation*}
where $\mathbf{C}_0 = \mathbf{A}_{0,0}\mathbf{c}_{0,0}+\mathbf{A}_{0,1}\mathbf{\Psi}_4\mathbf{c}_{0,1}+\mathbf{A}_{0,1}\mathbf{\Psi}_1\mathbf{c}_{1,0}+\mathbf{A}_{0,3}\mathbf{\Psi}_1\mathbf{c}_{3,0}+
\mathbf{A}_{0,4}\mathbf{c}_{4,0}+\mathbf{A}_{0,0}\mathbf{\Psi}_4\mathbf{c}_{4,4}$ and $\mathbf{C}_1 = \mathbf{A}_{1,0}\mathbf{c}_{0,0}+\mathbf{A}_{1,1}\mathbf{\Psi}_4\mathbf{c}_{0,1}+\mathbf{A}_{1,1}\mathbf{\Psi}_1\mathbf{c}_{1,0}+\mathbf{A}_{1,3}\mathbf{\Psi}_1\mathbf{c}_{3,0}+
\mathbf{A}_{1,4}\mathbf{c}_{4,0}+\mathbf{A}_{1,0}\mathbf{\Psi}_4\mathbf{c}_{4,4}$. %
Since the $2m \times 2m$ matrix $\left[ \begin{smallmatrix}
\mathbf{A}_{0,2} & \mathbf{A}_{0,3} \\
\mathbf{A}_{1,2} & \mathbf{A}_{1,3} 
\end{smallmatrix} \right]$ is full rank and $\mathbf{\Psi}_4$ is also full rank, the data chunks $\{ \mathbf{c}_{2,0}, \mathbf{c}_{2, 2}\}$ can be recovered from $\{ \mathbf{c}_{0, 0}, \mathbf{c}_{0, 1} \} \cup \{ \mathbf{c}_{1, 0} \} \cup \{ \mathbf{c}_{3, 0}
\} \cup \{\mathbf{c}_{4, 0}, \mathbf{c}_{4, 4} \}$. %
Similarly, 
\begin{itemize}
\item $\{ \mathbf{c}_{2, 1}, \mathbf{c}_{2, 3}\}$ can be recovered from $\{ \mathbf{c}_{0, 1} \} \cup \{\mathbf{c}_{1, 0}, \mathbf{c}_{1, 1} \} \cup  \{\mathbf{c}_{3, 1} \} \cup \{\mathbf{c}_{4, 1}, \mathbf{c}_{4, 5}\}$ using the second block rows of the two parity-check equations defined in  \eqref{eqn:def-C};
\item $\{ \mathbf{c}_{2, 4}, \mathbf{c}_{2, 6} \}$ can be recovered from $\{ \mathbf{c}_{0, 4}, \mathbf{c}_{0, 5} \} \cup \{\mathbf{c}_{1, 4} \} \cup \{\mathbf{c}_{3, 4} \} \cup \{ \mathbf{c}_{4,4}\}$ using the fifth block rows of the two parity-check equations defined in \eqref{eqn:def-C};
\item $\{ \mathbf{c}_{2, 5}, \mathbf{c}_{2, 7} \}$ can be recovered from $\{ \mathbf{c}_{0, 5} \} \cup \{ \mathbf{c}_{1, 4}, \mathbf{c}_{1, 5} \} \cup \{\mathbf{c}_{3, 5} \} \cup \{\mathbf{c}_{4, 5}\}$ using the sixth block rows of the two parity-check equations defined in \eqref{eqn:def-C}. %
\end{itemize}
We check that $\{ \mathbf{c}_{0,0}, \mathbf{c}_{0,1}\} \cup \{ \mathbf{c}_{0,1}\} \cup \{ \mathbf{c}_{0,4}, \mathbf{c}_{0,5}\} \cup \{ \mathbf{c}_{0,5}\} = \mathcal{M}_{0}^{(1,0)}$, $\{ \mathbf{c}_{1,0}\} \cup \{ \mathbf{c}_{1,0}, \mathbf{c}_{1,1}\} \cup \{ \mathbf{c}_{1,4}\} \cup \{ \mathbf{c}_{1,4}, \mathbf{c}_{1,5}\} = \mathcal{M}_{1}^{(1,0)}$, $\{ \mathbf{c}_{3,0}\} \cup \{ \mathbf{c}_{3,1}\} \cup \{ \mathbf{c}_{3,4}\} \cup \{ \mathbf{c}_{3,5}\} = \mathcal{M}_{3}^{(1,0)}$ and $\{ \mathbf{c}_{4,0}, \mathbf{c}_{4,4}\} \cup \{ \mathbf{c}_{4,1}, \mathbf{c}_{4,5}\} \cup \{ \mathbf{c}_{4,4}\} \cup \{ \mathbf{c}_{4,5}\} = \mathcal{M}_{4}^{(1,0)}$. %
Therefore, node $\mathbf{c}_2$ can be fully repaired by connecting to $d = 4$ helper nodes and accessing the data chunks in $\mathcal{M}^{(1, 0)} = \mathcal{M}_{0}^{(1,0)} \cup \mathcal{M}_{1}^{(1,0)} \cup \mathcal{M}_{3}^{(1,0)} \cup \mathcal{M}_{4}^{(1,0)}$. %
Each $\mathcal{M}_{i}^{(1,0)}$ for $i = 0, 1, 3 ,4$ contains exactly $4$ data chunks. %
Consequently, node $\mathbf{c}_2$ can be repaired by accessing a total of $16$ data chunks ($16m$ bits) from these sets, which achieves the optimal repair bandwidth lower bound given in \eqref{eqn:lower-bound}. %
The same argument holds for all other nodes of the $(5,3, 8m)$ code $\mathcal{C}_1$.
As a result, $\mathcal{C}_1$ achieves optimal access bandwidth for any single-node failure. %
\hfill $\blacksquare$
\end{example}

\section{Generic construction of binary MDS array codes with optimal repair bandwidth}
\label{sec:G-cons-optrepair}

In this section, stemming from Generic Construction \ref{G-construction1}, we propose another generic approach for constructing $(k+r, k, l)$ binary MDS array codes, which achieves optimal repair bandwidth for any single-node failure while requiring a smaller sub-packetization level $l$ than $\mathcal{C}_1$. %

\subsection{Generic construction}
Let $r > 2$ be an even number. %
Consider an arbitrary $(K+r, K, m)$ binary MDS array code with parity-check matrix $\mathbf{A}$ defined in \eqref{eqn:Block-PC-matrix-base}. 
Taking this $(K+r, K, m)$ binary MDS array code as the base code, we next construct a $(k+r, k, l)$ binary MDS array code with optimal repair bandwidth. To the best of our knowledge, the new code achieves the smallest sub-packetization among existing binary MDS codes with optimal repair bandwidth. %

\begin{G-construction}
\label{cons:G-cons-optrepair}
Consider integers $s$ and $k$ satisfying $s = \frac{r}{2}$, $k \leq \lfloor \frac{(s+1)(K+r)}{r} \rfloor - r$ and $s+1 \mid k+r$. %
Let $g = \frac{k+r}{s+1}$ and $l = ms^g$. %
Let $\mathcal{C}_2$ denote the $(k+r, k, l)$ binary array code defined by the parity-check matrix $\mathbf{H} = [\mathbf{H}_{i, j}]_{0 \leq i \leq r-1, 0 \leq j \leq k+r-1}$, in which every block entry $\mathbf{H}_{i, j}$ is an $l \times l$ matrix over $\mathbb{F}_2$ constructed based on $\mathbf{A}$ and arbitrarily given $m \times m$ full-rank matrices $\mathbf{\Psi}_1$, $\mathbf{\Psi}_2$, $\mathbf{\Psi}_3$ and $\mathbf{\Psi}_4$ (called coefficient matrices) such that the matrix $\left[ \begin{smallmatrix} \mathbf{\Psi}_1 & \mathbf{\Psi}_4 \\ \mathbf{\Psi}_3 & \mathbf{\Psi}_2 \end{smallmatrix} \right]$ is also full rank. %
Regard $\mathbf{H}_{i,j}$ as an $s^g \times s^g$ block matrix with every block entry an $m\times m$ matrix. %
For $0 \leq a, b \leq s^g-1$, the $(a, b)$-th block entry in $\mathbf{H}_{i,j}$, denoted by  $\mathbf{H}_{i, j}(a, b)$, is constructed as follows. 
Notice that for every $0 \leq j \leq k+r-1$, it can be expressed as $j = v(s+1)+u$ for all $0 \leq v \leq g-1$,  $0 \leq u \leq s$. 
\begin{itemize}
  \item For $j = v(s+1)+u$ with $0 \leq v \leq g-1$ and $0 \leq u \leq s-1$, 
\begin{equation}
\label{eqn: G-cons-optrepair}
\begin{aligned}
&\mathbf{H}_{i, j}(a, b) = \begin{cases}
\mathbf{A}_{i, 2j-2v-u}\mathbf{\Psi}_1 & a_v < u, ~b = a \\
\mathbf{A}_{i, 2j-2v-u} & a_v = u, ~b = a \\
\mathbf{A}_{i, 2j-2v-u}\mathbf{\Psi}_2 & a_v > u, ~b = a \\
\mathbf{A}_{i, 2(j-u-v)+w}\mathbf{\Psi}_3 & \begin{array}{@{}l@{}} a_v = u,~b = a(v,w),~0 \leq w < u \end{array} \\
\mathbf{A}_{i, 2(j-u-v)+ w}\mathbf{\Psi}_4 &   \begin{array}{@{}l@{}} a_v = u,~b = a(v,w),~u < w \leq s-1 \end{array} \\
\mathbf{0} & \text{otherwise} 
\end{cases},
\end{aligned}
\end{equation}
where $a_v$ represents the $v$-th $s$-ary symbol in the $s$-ary expression $\mathbf{s}_a = [a_{g-1} ~ a_{g-2} ~ \cdots ~ a_0 ]$ of $a$, and $a(v,w)$ represents the integer smaller than $s^{\frac{k+r}{s+1}}$ with $\mathbf{s}_{a(v,w)} = [a_{g-1} ~ \ldots a_{v+1} ~w ~ a_{v-1}~ \ldots~a_0 ]$. 
\item For $j = (v+1)(s+1)-1$ with $0 \leq v \leq g-1$, 
\begin{equation}
\label{eqn:G-cons-optrepair-1}
\mathbf{H}_{i, j} = \mathbf{I}_{s^{g-v-1}} \otimes \bar{\mathbf{A}}_{i, j},
\end{equation}
where $\bar{\mathbf{A}}_{i, j}$ denotes the $s \times s$ block diagonal matrix with block diagonal entries equal to $\mathbf{I}_{s^{v}} \otimes \mathbf{A}_{i, 2(j-v)-s}, \mathbf{I}_{s^{v}} \otimes \mathbf{A}_{i, 2(j-v)-(s-1)}, \cdots, \mathbf{I}_{s^{v}} \otimes \mathbf{A}_{i, 2(j-v)-1}$, \emph{i.e.},
\begin{equation*}
\bar{\mathbf{A}}_{i, j} = {\rm diag}(\mathbf{I}_{s^{v}} \otimes \mathbf{A}_{i, 2(j-v)-s}, \cdots, \mathbf{I}_{s^{v}} \otimes \mathbf{A}_{i, 2(j-v)-1}). %
\end{equation*} 
\hfill $\blacksquare$
\end{itemize}
\end{G-construction}

The $k+r$ nodes of $\mathcal{C}_2$ constructed by Generic Construction \ref{cons:G-cons-optrepair} are divided into $g$ groups of size $s+1$ (while in $\mathcal{C}_1$ the size of each group is at most $s$). %
One can easily observe that for code $\mathcal{C}_2$, every block entry $\mathbf{H}_{i,j}$ of $\mathbf{H}$ in each group exhibits the following characteristic: for $0 \leq u \leq s-1$ (as defined in \eqref{eqn: G-cons-optrepair}),  it follows the same pattern as in code $\mathcal{C}_1$ (see \eqref{eqn: G-construction2}), while for $u = s$ (as defined in \eqref{eqn:G-cons-optrepair-1}), it is an $s^g \times s^g$ block diagonal matrix with every block diagonal entry an $m \times m$ matrix. %
Similar to Generic Construction \ref{G-construction1}, a simplified yet valid instantiation of Generic Construction \ref{cons:G-cons-optrepair} can be obtained by setting $\mathbf{\Psi}_1 = \mathbf{\Psi}_2 = \mathbf{\Psi}_3 = \mathbf{I}_m$ and choosing $\mathbf{\Psi}_4$ as an arbitrary full-rank matrix such that $\mathbf{I}_m + \mathbf{\Psi}_4$ is also full rank. %
We demonstrate in the following that code $\mathcal{C}_2$ possesses both the MDS property and the optimal repair property, where the latter means that any failed node can be repaired by connecting to $d = k+s-1$ helper nodes and by downloading $\frac{l}{s}$ bits from each helper node. %

\begin{theorem}
\label{theorem:MDS-2}
In Generic Construction \ref{cons:G-cons-optrepair}, if the $(K+r, K, m)$ base code is MDS, then the $(k+r, k, ms^{\frac{k+r}{s+1}})$ binary array code $\mathcal{C}_2$ is MDS.
\begin{IEEEproof}
Please refer to Appendix-\ref{appendix:proof-MDS-2}. 
\end{IEEEproof}
\end{theorem}

\begin{theorem}
\label{theorem:optimal-repair}
In Generic Construction \ref{cons:G-cons-optrepair}, if the $(K+r, K, m)$ base code is MDS, then the $(k+r, k, ms^{\frac{k+r}{s+1}})$ binary array code $\mathcal{C}_2$ achieves optimal repair bandwidth with $d = k+s-1$ helper nodes. %
Specifically, the repair of any single node among the first $s$ nodes in each group possesses the optimal access property. %
\end{theorem}

The proof of Theorem \ref{theorem:optimal-repair} will be given in the next subsection. 
To the best of our knowledge, the only known constructions of binary MDS array codes with optimal repair bandwidth and with $r \geq 4$ are those presented in  \cite{Lei-TCOM24-Binary-optimal}\cite{Hou-25TIT-Generic-optimal-repair} with sub-packetization $l = ms^{k+r}$, in \cite{Hou-20TIT-transform} with $l = ms^{\lceil \frac{k}{s}\rceil+\lceil \frac{r}{s}\rceil}$, and in \cite{Tang-22TCOM-transform} with $l = mr^{\lceil \frac{k+r}{r}\rceil}$. %
Compared to these constructions, code $\mathcal{C}_1$ achieves reduced sub-packetization $ms^{\lceil \frac{k+r}{s} \rceil}$, while code $\mathcal{C}_2$ attains even lower sub-packetization $ms^{\frac{k+r}{s+1}}$, a value that is the smallest for the class of binary MDS array codes with optimal repair bandwidth known to date. %

For binary MDS array codes with optimal repair bandwidth, keeping the amount of data bits accessed during node repair as small as possible is also of significant practical importance, as it directly reduces I/O overhead and improves repair efficiency. %
Although the binary MDS array codes presented in \cite{Lei-TCOM24-Binary-optimal} \cite{Hou-25TIT-Generic-optimal-repair} achieve the optimal repair bandwidth $\frac{dl}{s}$, the repair of any failed node requires accessing all $l$ bits from each helper node. %
Consequently, the amount of the data bits accessed for repairing a failed node is $dl$, which is exactly $s$ times the optimal one. %
In contrast, as to be clear in the discussion in the next subsection, there are $\frac{s}{s+1}(k+r)$ nodes in $\mathcal{C}_2$ such that the repair of each of them has the optimal access property, whereas the repair of each of the remaining $\frac{1}{s+1}(k+r)$ nodes requires accessing all $l$ bits from each helper node. %
It turns out that in order to achieve optimal repair bandwidth $\frac{dl}{s}$, the average number of data bits accessed for repairing any single-node failure is $\frac{2s}{s+1}\frac{dl}{s}$, less than twice the optimal one. 
To the best of our knowledge, $\mathcal{C}_2$ has the smallest average number of data bits accessed for repairing any single-node failure among existing binary MDS array codes with optimal repair but not optimal access bandwidth. %

\subsection{Justification of the optimal repair property of $\mathcal{C}_2$}
Consider the $(k+r, k, l = ms^{\frac{k+r}{s+1}})$ binary MDS array code $\mathcal{C}_2$ constructed by Generic Construction \ref{cons:G-cons-optrepair}. %
This subsection will justify the optimal repair property of $\mathcal{C}_2$, so that Theorem \ref{theorem:optimal-repair} can be proved. %
Code $\mathcal{C}_2$ is said to have the optimal repair property if any failed node can be repaired by connecting to $d = k+s-1$ helper nodes and downloading exactly $\frac{l}{s}$ bits from each helper node. %
In the following, we prove that any failed node of $\mathcal{C}_2$ can be recovered by downloading exactly $\frac{l}{s}$ bits from each of the $d$ helper nodes, which in turn implies that $\mathcal{C}_2$ possesses the optimal repair property. %
Moreover, if the failed node belongs to the first $s$ nodes within its group, then the accessed data bits from each helper node coincide with the downloaded data bits, thereby ensuring that the repair of this node satisfies the optimal access property. %
Throughout this section, unless otherwise specified, $u$ is an arbitrary integer between $0$ and $s$, and $v$ is an arbitrary integer between $0$ and $g-1$. %
We shall specify the amount of data bits downloaded from all helper nodes during the repair of node $\mathbf{c}_j$ with $j = v(s+1)+u$. %

Let $l' = s^{\frac{k+r}{s+1}}$. %
For each $0 \leq t \leq k+r-1$, represent the $t$-th node $\mathbf{c}_t$ as $\mathbf{c}_t = [\mathbf{c}_{t,0}^{\rm T} ~ \mathbf{c}_{t,1}^{\rm T} ~\cdots~ \mathbf{c}_{t,l'-1}^{\rm T}]^{\rm T}$, where each data chunk $\mathbf{c}_{t, i}$ is an $m$-bit column vector, and define the following data chunk set of the $t$-th node:
\begin{equation}
\label{eqn:chunk-opt-repair}
\mathcal{M}_t^{(v, u)} = 
\begin{cases}
\{ \mathbf{c}_{t, a}:~ 0 \leq a \leq l'-1, ~a_v = u \} & 0 \leq u \leq s-1 \\
\{ \sum\nolimits_{i=0}^{s-1}\mathbf{c}_{t,a+is^{v}}: 0 \leq a \leq l'-1, a_{v} = 0 \}  & u = s
\end{cases}, 
\end{equation}
where $a_v$ represents the $v$-th $s$-ary symbol in the $s$-ary expression $\mathbf{s}_a = [a_{g-1} ~ a_{g-2} ~ \cdots ~ a_0 ]$ of $a$. %

Let $\mathcal{N}^{(v)} = \{ v(s+1)+w: w \in \{ 0, 1, \cdots, s-1\}\}$ and $\bar{\mathcal{N}}^{(v)} = \{ 0, 1, \cdots, k+r-1 \} \setminus \mathcal{N}^{(v)}$. %
Further, for $0 \leq u \leq s-1$, corresponding to every subset $\mathcal{G}^{(v)} \subset \bar{\mathcal{N}}^{(v)}$ with cardinality $|\mathcal{G}^{(v)}| = k$, define $\bar{\mathcal{G}}^{(v,u)}$ as 
\begin{equation}
\label{eqn: def_idx_G-optrepair}
\bar{\mathcal{G}}^{(v,u)} = \mathcal{G}^{(v)} \cup (\mathcal{N}^{(v)} \setminus \{ v(s+1)+u \} ). %
\end{equation}
For $u = s$, define $\bar{\mathcal{G}}^{(v,s)}$ as
\begin{equation}
\label{eqn: def_idx_G-optrepair-1}
\bar{\mathcal{G}}^{(v,s)} = \bar{\mathcal{N}}^{(v)} \setminus \{ v(s+1)+s \}. 
\end{equation}

In order to repair any failed node $\mathbf{c}_j$ with $j = v(s+1)+u$ of $\mathcal{C}_2$, it is necessary to connect to $d = k+s-1$ helper nodes and download exactly $\frac{l}{s}$ bits from each helper node. %
Specifically, when $0 \leq u \leq s-1$, node $\mathbf{c}_{v(s+1)+u}$ can be repaired by connecting to the $s-1$ nodes in the set $\{ \mathbf{c}_i~:~i \in \mathcal{N}^{(v)} \setminus \{v(s+1)+u\}\}$ and arbitrary $k$ nodes from the set $\{ \mathbf{c}_i~:~i \in \bar{\mathcal{N}}^{(v)}\}$. %
In other words, the repair can be accomplished by connecting to all $d$ nodes in the set $\{ \mathbf{c}_i~:~i \in \bar{\mathcal{G}}^{(v,u)}\}$. %
When $u = s$, node $\mathbf{c}_{v(s+1)+s}$ can be repaired by connecting to all $d = k+s-1$ nodes from other groups, meaning that every node in the set $\{ \mathbf{c}_i~:~i \in \bar{\mathcal{G}}^{(v,s)}\}$ participates in the repair process of node $\mathbf{c}_{v(s+1)+s}$. %
Furthermore, during the repair of node $\mathbf{c}_{v(s+1)+u}$, the $\frac{l'}{s}$ data chunks downloaded from any helper node $\mathbf{c}_i$, $i \in \bar{\mathcal{G}}^{(v,u)}$, correspond exactly to the set $\mathcal{M}_i^{(v,u)}$ defined in \eqref{eqn:chunk-opt-repair}, and these data chunks collectively contain exactly $\frac{l}{s}$ bits. %
Consequently, the following theorem specifies the total amount of data downloaded from all helper nodes during the repair of node $\mathbf{c}_j$. %

\begin{theorem}
\label{theorem:optrepair-M}
For the code $\mathcal{C}_2$ constructed by Generic Construction \ref{cons:G-cons-optrepair}, node $\mathbf{c}_{j}$ with $j = v(s+1)+u$ can be recovered by downloading the $\frac{dl'}{s}$ data chunks in the set
\begin{equation}
\label{eqn:optrepair_M_1}
\mathcal{M}^{(v, u)} = \bigcup\nolimits_{i \in \bar{\mathcal{G}}^{(v,u)}} \mathcal{M}_i^{(v, u)}.  %
\end{equation}
\begin{IEEEproof}
Please refer to Appendix-\ref{appendix:proof-optrepair-M}. 
\end{IEEEproof}
\end{theorem}

The optimal repair property of $\mathcal{C}_2$, as justified by Theorem \ref{theorem:optrepair-M}, refers to the fact that for any single-node failure, the repair process requires connecting to $d = k+s-1$ helper nodes and downloading exactly $\frac{l}{s}$ bits from each helper node. %
One can verify that for $0 \leq u \leq s-1$, during the repair of node $\mathbf{c}_{v(s+1)+u}$, the set $\mathcal{M}_i^{(v, u)}$ comprising the $\frac{l'}{s}$ data chunks downloaded from each helper node $\mathbf{c}_i$, $i \in \bar{\mathcal{G}}^{(v,u)}$, is exactly the set of data chunks that need to be accessed. %
Consequently, the repair of each of the $\frac{s}{s+1}(k+r)$ nodes in the set $\{ \mathbf{c}_{v(s+1)+u}~:~0 \leq v \leq g-1, 0 \leq u \leq s-1 \}$ achieves the optimal access property. %
The total access volume for repairing these $\frac{s}{s+1}(k+r)$ nodes amounts to $\frac{s}{s+1}(k+r) \cdot \frac{dl}{s}$ bits. %
In contrast, for $u = s$, while the repair of node $\mathbf{c}_{v(s+1)+s}$ still requires downloading $\frac{l'}{s}$ data chunks from each helper node $\mathbf{c}_i$, $i \in \bar{\mathcal{G}}^{(v,s)}$, each of these downloaded data chunks is actually a linear summation of $s$ distinct data chunks, thus requiring the repair process to access all $l'$ data chunks from each helper node. %
As a result, the total access volume for repairing these $\frac{1}{s+1}(k+r)$ nodes amounts to $\frac{1}{s+1}(k+r) \cdot dl$ bits.
The above analysis leads to the following proposition. %

\begin{proposition}
\label{prop:average_number-access}
For the code $\mathcal{C}_2$ constructed by Generic Construction \ref{cons:G-cons-optrepair}, in order to achieve the optimal repair bandwidth $\gamma = \frac{dl}{s}$, the average number of data bits accessed for repairing any single-node failure is $\frac{2s}{s+1}\gamma$, less than twice the optimal one. %
\end{proposition}

\section{Conclusion}
\label{sec:Consclusion}
In this paper, given an arbitrary binary MDS array code with sub-packetization $m$ as the base code, we propose two generic approaches for constructing binary MDS array codes with optimal repair bandwidth or optimal access bandwidth for single-node failures. %
For every $s \leq r$, a $(k+r, k, ms^{\lceil \frac{k+r}{s}\rceil})$ code $\mathcal{C}_1$ with optimal access bandwidth can be constructed by Generic Construction \ref{G-construction1}.
The repair of a failed node of $\mathcal{C}_1$ requires connecting to $d = k+s-1$ helper nodes, in which $s-1$ helpers are designated and $k$ are free to select. %
The architectural differences of the parity-check matrices between $\mathcal{C}_1$ and previously proposed codes with the smallest sub-packetization are illustrated. %
Moreover, $\mathcal{C}_1$ generally achieves smaller sub-packetization and provides greater flexibility in the selection of its coefficient matrices. %
For even $r \geq 4$ and $s = \frac{r}{2}$ such that $s+1$ divides $k+r$, a $(k+r, k, ms^{\frac{k+r}{s+1}})$ code $\mathcal{C}_2$ with optimal repair bandwidth can be constructed by Generic Construction \ref{cons:G-cons-optrepair}, with $\frac{s}{s+1}(k+r)$ out of $k+r$ nodes having the optimal access property. %
In order to achieve the optimal repair bandwidth, the average number of data bits accessed for repairing any single-node failure of $\mathcal{C}_2$ is $\frac{2s}{s+1}\frac{dl}{d-k+1}$, less than twice the optimal one.
To the best of our knowledge, $\mathcal{C}_2$ possesses the smallest sub-packetization among existing binary MDS array codes with optimal repair bandwidth known to date, and among existing binary MDS array codes with optimal repair but not optimal access bandwidth, $\mathcal{C}_2$ has the smallest average number of data bits accessed for repairing any single-node failure. %
As a future work, it is interesting to further extend Generic Construction \ref{G-construction1} to construct binary MDS array codes with optimal access bandwidth that allow arbitrary selection of all $d$ helper nodes. %
It is also of interest to extend Generic Construction \ref{cons:G-cons-optrepair} to be applicable to every $s \leq r$. %

{\appendix}

\subsection{Proof of Lemma \ref{lemma:U=r}}
\label{appendix:lemma-U=r}
According to \eqref{eqn: G-construction2} and \eqref{eqn: G-construction2-1}, for each $0 \leq j \leq k+r-1$, every block diagonal entry of the block matrix $\mathbf{H}_{i,j}$ is either $\mathbf{A}_{i, j}$, $\mathbf{A}_{i, j}\mathbf{\Psi}_1$, or $\mathbf{A}_{i, j}\mathbf{\Psi}_2$. %
Consequently, for each $0 \leq j \leq k+r-1$, every block row of $\mathbf{H}_{i, j}$ contains one of these matrices. %
It follows that for each $j \in \{ j_0, \ldots, j_{r-1}\}$, $\mathbf{H}''^{(a)}$, $0 \leq a \leq l'-1$, contains a nonzero block column belonging to $\{\mathbf{L}_{j}, \mathbf{L}_{j}\mathbf{\Psi}_1, \mathbf{L}_{j}\mathbf{\Psi}_2 \}$. %
This implies that for all $0 \leq a \leq l'-1$, we have $\{ j_0, \ldots, j_{r-1}\} \subseteq \mathcal{U}^{(a)}$, and consequently, $\min|\mathcal{U}^{(a)}| \geq r$. %
To prove the equality, we show that there exists an index $a \in \{ 0, 1, \ldots, l'-1 \}$ such that $\mathcal{U}^{(a)} = \{ j_0, j_1, \ldots, j_{r-1}\}$, which is equivalent to prove that for some $a \in \{ 0, 1, \ldots, l'-1 \}$ and for each $j \in \{j_0, j_1,\ldots, j_{r-1}\}$, every nonzero block entry in the $a$-th block row of $\mathbf{H}_{i, j}$ belongs to the set
\begin{equation}
\label{eqn:appen-proof-MDS2-0}
\{ \mathbf{A}_{i,j_0}\mathbf{\Psi}, \mathbf{A}_{i,j_1}\mathbf{\Psi}, \ldots, \mathbf{A}_{i,j_{r-1}}\mathbf{\Psi}~:~\mathbf{\Psi} \in \{\mathbf{I}_m, \mathbf{\Psi}_1, \mathbf{\Psi}_2, \mathbf{\Psi}_3, \mathbf{\Psi}_4 \} \}. 
\end{equation}

Define a (possibly empty) set $\mathcal{K}$ as
\begin{equation}
\label{eqn:appen-proof-MDS2-1}
\begin{aligned}
\mathcal{K} = &\{ v: 0 \leq v \leq g, \{ vs, vs+1, \ldots, vs+s-1\} \subseteq \{ j_0, j_1, \ldots, j_{r-1}\}\}. 
\end{aligned}
\end{equation}
Clearly, one can easily verify that there always exists an $a \in \{ 0, 1, \ldots, l'-1\}$ such that if $v \in \{0, 1, \ldots, g \} \backslash \mathcal{K}$ then 
\begin{equation}
\label{eqn:appen-proof-MDS2-2}
\begin{aligned}
&vs+a_v \notin \{ j_0, j_1, \ldots, j_{r-1}\}, 
\end{aligned}
\end{equation}
where $0 \leq a_v \leq s-1$. %
With such an $a$ chosen, our goal is to justify that for any $j \in \{j_0, j_1, \ldots, j_{r-1} \}$, every nonzero block entry in the $a$-th block row of $\mathbf{H}_{i, j}$ indeed belongs to the set given in \eqref{eqn:appen-proof-MDS2-0}. %

Recall that each $j \in \{ j_0, j_1, \ldots, j_{r-1}\}$ can be expressed as $j = vs+u$ for some $0 \leq v \leq g$, $0 \leq u \leq s-1$. %
We consider the following two cases. %
\begin{itemize}
\item ($v \in \{0, 1, \ldots, g \} \backslash \mathcal{K}$): Based on \eqref{eqn:appen-proof-MDS2-2}, the condition $vs+a_v \neq j$ implies that $a_v \neq u$. %
  According to \eqref{eqn: G-construction2} and \eqref{eqn: G-construction2-1}, the $a$-th block row of $\mathbf{H}_{i, j}$ contains a single nonzero block entry, either $\mathbf{A}_{i, j}\mathbf{\Psi}_1$ or $\mathbf{A}_{i, j}\mathbf{\Psi}_2$, which belongs to the set given in \eqref{eqn:appen-proof-MDS2-0}. 
\item ($v \in \mathcal{K}$): If $a_v \neq u$, then the same argument applies as above. %
  And if $a_v = u$, then the $a$-th block row of $\mathbf{H}_{i, j}$ contains $s$ nonzero block entries. %
   Specifically, when $0 \leq w < u$, the $(a, a(v, w))$-th block entry of $\mathbf{H}_{i, j}$ is $\mathbf{A}_{i, vs+w}\mathbf{\Psi}_3$; when $w = u$, the $(a, a(v, w))$-th block entry of $\mathbf{H}_{i, j}$ is $\mathbf{A}_{i, vs+u}$; and when $u < w \leq s-1$, the $(a, a(v, w))$-th block entry of $\mathbf{H}_{i, j}$ is $\mathbf{A}_{i, vs+w}\mathbf{\Psi}_4$. %
  Based on \eqref{eqn:appen-proof-MDS2-1}, they all belong to the set given in \eqref{eqn:appen-proof-MDS2-0}. %
\end{itemize}
Thus, we have proved that \eqref{eqn:appen-proof-MDS2-00} holds. %
The proof is complete. %

\subsection{Proof of Lemma \ref{lemma:J_a}}
\label{appendix:lemma-J_a}
We will argue by induction on the cardinality of the set $\mathcal{U}^{(a)}$ to prove this lemma. %
By Lemma \ref{lemma:U=r}, to establish the induction basis, we first prove that this lemma holds for all $a$ such that $|\mathcal{U}^{(a)}| = r$. %
Let $a$ be one of the values that satisfy $|\mathcal{U}^{(a)}| = r$. %
For the given $a$, all the nonzero block columns of $\mathbf{H}''^{(a)}$ belong to the set
\begin{equation*}
\{ \mathbf{L}_{j_0}\mathbf{\Psi}, \mathbf{L}_{j_1}\mathbf{\Psi}, \ldots, \mathbf{L}_{j_{r-1}}\mathbf{\Psi}: \mathbf{\Psi} \in \{\mathbf{I}_m, \mathbf{\Psi}_1, \mathbf{\Psi}_2, \mathbf{\Psi}_3, \mathbf{\Psi}_4 \} \}.
\end{equation*}
Thus, $\mathbf{H}''^{(a)}[\mathbf{x}_0^{\rm T}~\mathbf{x}_1^{\rm T}~\ldots~\mathbf{x}_{rl'-1}^{\rm T}]^{\rm T} = \mathbf{0}$ implies  
\begin{equation}
[\mathbf{L}_{j_0}~\mathbf{L}_{j_1}~\ldots~\mathbf{L}_{j_{r-1}}] [\bar{\mathbf{x}}_{j_0}^{\rm T}~\bar{\mathbf{x}}_{j_1}^{\rm T}~\ldots~\bar{\mathbf{x}}_{j_{r-1}}^{\rm T}]^{\rm T} = \mathbf{0}. %
\end{equation}
Recall from \eqref{eqn:J_a_partition} that $\bigcup_{0 \leq h  \leq r-1} \mathcal{J}^{(a)}(j_h) = \mathcal{J}^{(a)}$, where each $\mathcal{J}^{(a)}(j_h)$ contains either one or two indices. %
Consequently, for $0 \leq i \leq r-1$ 
\begin{itemize}
\item if $|\mathcal{J}^{(a)}(j_i)| = 1$ with $\mathcal{J}^{(a)}(j_i) = \{ t \}$, then $\bar{\mathbf{x}}_{j_i} = \mathbf{x}_t$;
\item if $|\mathcal{J}^{(a)}(j_i)| = 2$ with $\mathcal{J}^{(a)}(j_i) = \{t, t'\}$, then $\bar{\mathbf{x}}_{j_i}$ is a linear combination of $\mathbf{x}_t$ and $\mathbf{x}_{t'}$. %
\end{itemize}
The block columns $\mathbf{L}_{j_0}, \mathbf{L}_{j_1}, \ldots, \mathbf{L}_{j_{r-1}}$ form the block sub-matrix corresponding to block columns $j_0, j_1, \ldots, j_{r-1}$ of the parity-check matrix $\mathbf{A}$ of the $(K+r, K, m)$ base code. %
Given that the base code is MDS, the matrix $[\mathbf{L}_{j_0}~\mathbf{L}_{j_1}~\ldots~\mathbf{L}_{j_{r-1}}]$ is full-rank. %
It therefore follows that $\bar{\mathbf{x}}_{j_i} = \mathbf{0}$ for all $0 \leq i \leq r-1$. %
In the following, we determine the specific indices $t$ involved in each case and show that $\mathbf{x}_t = \mathbf{0}$ for all $t \in \mathcal{J}^{(a)}$. 

Based on \eqref{eqn: G-construction2} and \eqref{eqn: G-construction2-1}, for each $0 \leq j \leq k+r-1$, all nonzero block entries of block matrix $\mathbf{H}_{i,j}$ belong to the set \scalebox{0.9}{$\{ \mathbf{A}_{i, j-u}\mathbf{\Psi}_3, \mathbf{A}_{i, j-u+1}\mathbf{\Psi}_3, \ldots ,\mathbf{A}_{i, j-1}\mathbf{\Psi}_3, \mathbf{A}_{i, j}\mathbf{\Psi}_1, \mathbf{A}_{i, j}, \mathbf{A}_{i, j}\mathbf{\Psi}_2, \mathbf{A}_{i, \sigma(j+1)}\mathbf{\Psi}_4, \mathbf{A}_{i, \sigma(j+2)}\mathbf{\Psi}_4, \ldots, \mathbf{A}_{i, \sigma(j-u+s-1)}\mathbf{\Psi}_4 \}$} . %
One can verify that if $a_v \neq u$, then the $a$-th block row of $\mathbf{H}_{i,j}$ contains a single nonzero block entry,  either $\mathbf{A}_{i, j}\mathbf{\Psi}_1$ or $\mathbf{A}_{i, j}\mathbf{\Psi}_2$, located in the $a$-th block column. %
Conversely, if $a_v = u$, the $a$-th block row of $\mathbf{H}_{i,j}$ contains exactly $s$ nonzero block entries. %
In particular, when $0 \leq w < u$, $\mathbf{A}_{i, j-u+w}\mathbf{\Psi}_3$ only appears in the $a$-th block row and $a(v, w)$-th block column of $\mathbf{H}_{i, j}$; when $w = u$, $\mathbf{A}_{i, j}$ only appears in the $a$-th block row and $a$-th block column of $\mathbf{H}_{i, j}$; and when $u < w \leq s-1$, $\mathbf{A}_{i, \sigma(j-u+w)}\mathbf{\Psi}_4$ only appears in the $a$-th block row and $a(v, w)$-th block column of $\mathbf{H}_{i, j}$. %
Recall that for each $0 \leq h \leq r-1$, the index $j_h$ can be expressed as $j_h = v_hs+u_h$ for some $0 \leq v_h \leq g$, $0 \leq u_h \leq s-1$. %
According to the above characteristic of $\mathbf{H}_{i, j}$, we partition $\{0, 1, \ldots, r-1\}$ into three disjoint subsets $\mathcal{W}_1$, $\mathcal{W}_2$, $\mathcal{W}_3$ as follows:
\begin{equation}
\label{eqn:W-pert}
\begin{aligned}
& \mathcal{W}_1 = \{h: 0 \leq h \leq r-1, a_{v_h} = u_h \} \\
&~~~~~~ \cup \{ h: 0 \leq h \leq r-1, \nexists p \in \{0, 1, \ldots, r-1\} ~{\rm s.t.}~ v_p = v_h ~{\rm and}~ u_p = a_{v_h}\}, \\
& \mathcal{W}_2 = \{h: 0 \leq h \leq r-1, a_{v_h} > u_h \} \\
&~~~~~~ \cap \{ h: 0 \leq h \leq r-1, \exists p \in \{0, 1, \ldots, r-1\} ~{\rm s.t.}~ v_p = v_h ~{\rm and}~ u_p = a_{v_h}\}, \\
& \mathcal{W}_3 = \{h: 0 \leq h \leq r-1, a_{v_h} < u_h \} \\
&~~~~~~ \cap \{ h: 0 \leq h \leq r-1, \exists p \in \{0, 1, \ldots, r-1\} ~{\rm s.t.}~ v_p = v_h ~{\rm and}~ u_p = a_{v_h}\}. \\
\end{aligned}
\end{equation}

It is clear that for each $h \in \mathcal{W}_1$, we have $\mathcal{J}^{(a)}(j_h) = \{ hl'+a\}$. %
Thus, we conclude that 
\begin{equation}
\label{eqn:W-1-MDS1}
\bar{\mathbf{x}}_{j_h} = \mathbf{x}_{hl'+a} = \mathbf{0}
\end{equation}
for all $h \in \mathcal{W}_1$. %

Considering the case $h \in \mathcal{W}_2$, observe that $\mathcal{J}^{(a)}(j_h) = \{ hl'+a, pl'+a(v_h, u_h)\}$ which implies that 
\begin{equation}
\label{eqn:append-proof-MDS2-W2}
\bar{\mathbf{x}}_{j_h} = \mathbf{\Psi}_2\mathbf{x}_{hl'+a} + \mathbf{\Psi}_3\mathbf{x}_{pl'+a(v_h, u_h)} = \mathbf{0}.
\end{equation}
Let us consider $\mathbf{H}''^{(a(v_h, u_h))}$, whose nonzero block columns are described as follows:
\begin{equation*}
\begin{aligned}
&\mathcal{U}^{(a(v_h, u_h))} = \{ j_0, j_1, \ldots, j_{r-1} \}, \\
&\mathcal{J}^{(a(v_h, u_h))}(j_p) = \{ hl'+a, pl'+a(v_h, u_h) \}.
\end{aligned}
\end{equation*}
From this we obtain that
\begin{equation}
\label{eqn:append-proof-MDS2-W3}
\mathbf{\Psi}_4\mathbf{x}_{hl'+a} + \mathbf{\Psi}_1\mathbf{x}_{pl'+a(v_h, u_h)} = \mathbf{0}.
\end{equation}
Recall that $\left[ \begin{smallmatrix} \mathbf{\Psi}_1 & \mathbf{\Psi}_4 \\ \mathbf{\Psi}_3 & \mathbf{\Psi}_2 \end{smallmatrix} \right]$ has full rank $2m$, \eqref{eqn:append-proof-MDS2-W2} and \eqref{eqn:append-proof-MDS2-W3} imply that $\mathbf{x}_{hl'+a} = \mathbf{x}_{pl'+a(v_h, u_h)} = \mathbf{0}$. %

For $h \in \mathcal{W}_3$, it is very similar to $\mathcal{W}_2$. %
One can observe that $\mathcal{J}^{(a)}(j_h) = \{ hl'+a, pl'+a(v_h,u_h)\}$ which implies that
\begin{equation}
\label{eqn:m_sec-1}
\bar{\mathbf{x}}_{j_h} = \mathbf{\Psi}_1\mathbf{x}_{hl'+a} + \mathbf{\Psi}_4\mathbf{x}_{pl'+a(v_h, u_h)} = \mathbf{0}. %
\end{equation}
Let us consider $\mathbf{H}''^{(a(v_h, u_h))}$, whose nonzero block columns are described as follows:
\begin{equation*}
\begin{aligned}
&\mathcal{U}^{(a(v_h, u_h))} = \{ j_0, j_1, \ldots, j_{r-1} \}, \\
&\mathcal{J}^{(a(v_h, u_h))}(j_p) = \{ hl'+a, pl'+a(v_h, u_h) \}.
\end{aligned}
\end{equation*}
From this we obtain that
\begin{equation}
\label{eqn:m_sec-2}
\mathbf{\Psi}_3\mathbf{x}_{hl'+a} + \mathbf{\Psi}_2\mathbf{x}_{pl'+a(v_h, u_h)} = \mathbf{0}.
\end{equation}
From \eqref{eqn:m_sec-1} and \eqref{eqn:m_sec-2}, one can readily compute that $\mathbf{x}_{hl'+a} = \mathbf{x}_{pl'+a(v_h, u_h)} = \mathbf{0}$. %
We obtain that $\mathbf{x}_t = \mathbf{0}$ for all $t \in \mathcal{J}^{(a)}(j_h)$. %
This completes the induction basis, proving $\mathbf{x}_t = \mathbf{0}$ for all  $t \in \mathcal{J}^{(a)}$ when $|\mathcal{U}^{(a)}| = r$.

In what follows, assume $\mathbf{x}_t = \mathbf{0}$ for all  $t \in \mathcal{J}^{(a)}$ and all $a$ with $|\mathcal{U}^{(a)}| \leq w - 1$, where $w \geq r$. %
We will prove that for all $a$ such that $|\mathcal{U}^{(a)}| = w$, $\mathbf{x}_t = \mathbf{0}$ for all $t \in \mathcal{J}^{(a)}$. %
According to \eqref{eqn:J_a_partition}, it is equivalent to prove that for every $j \in \mathcal{U}^{(a)}$, $\mathbf{x}_t = \mathbf{0}$ for all $t \in \mathcal{J}^{(a)}(j)$. %
For any $j \in \mathcal{U}^{(a)} \backslash \{ j_0, \ldots, j_{r-1} \}$, there exists a unique $h \in \{ 0, 1, \ldots, r-1\}$ such that $a_{v_h} = u_h$ and $0 \leq j-v_hs \leq s-1$, with the corresponding set $\mathcal{J}^{(a)}(j) = \{ hl'+a(v_h, \alpha)\}$, where $\alpha = j-v_hs$. %
Consider the matrix $\mathbf{H}''^{(a(v_h, \alpha))}$. %
We observe that for the chosen $j$, no $p \in \{0, 1, \ldots, r-1 \}$ simultaneously satisfies $v_p = v_h$ and $u_p = \alpha$. %
Thus, we have $\mathcal{U}^{(a(v_h, \alpha))} \subset \mathcal{U}^{(a)}$ and $j \notin \mathcal{U}^{(a(v_h, \alpha))}$, which implies $|\mathcal{U}^{(a(v_h, \alpha))}| \leq w-1$. %
We can conclude that the induction hypothesis applies and $\mathbf{x}_t = \mathbf{0}$ for all $t \in \mathcal{J}^{(a(v_h, \alpha))}$. %
Furthermore, $\mathbf{x}_{hl'+a(v_h, \alpha)} = \mathbf{0}$ since $hl'+a(v_h, \alpha)$ belongs to  $\mathcal{J}^{(a(v_h, \alpha))}$. %
Rephrasing this, we have established that for every $j \in \mathcal{U}^{(a)} \backslash \{ j_0, \ldots, j_{r-1} \}$, $\mathbf{x}_t = \mathbf{0}$ for all $t \in \mathcal{J}^{(a)}(j)$. %

Finally, we consider the variables $\mathbf{x}_t$ for all $t \in \mathcal{J}^{(a)}(j_h)$, where $0 \leq h \leq r-1$. %
Note that $\mathbf{H}''^{(a)}$ reduces to containing only the nonzero block columns from the set $\{ \mathbf{L}_{j_0}\mathbf{\Psi}, \mathbf{L}_{j_1}\mathbf{\Psi}, \ldots, \mathbf{L}_{j_{r-1}}\mathbf{\Psi}~:~\mathbf{\Psi} \in \{ \mathbf{I}_m, \mathbf{\Psi}_1, \mathbf{\Psi}_2, \mathbf{\Psi}_3, \mathbf{\Psi}_4 \} \}$. %
We directly conclude that for every $0 \leq h \leq r-1$, $\mathbf{x}_t = \mathbf{0}$ for all $t \in \mathcal{J}^{(a)}(j_h)$ due to the MDS property of the $(K+r, K, m)$ base code. %
Since the proof of this claim follows exactly the same reasoning as the induction basis proof above, we omit here. %
Consequently, for each $0 \leq a \leq l'-1$, $\mathbf{H}''^{(a)}[\mathbf{x}_0^{\rm T}~\mathbf{x}_1^{\rm T}~\ldots~\mathbf{x}_{rl'-1}^{\rm T}]^{\rm T} = \mathbf{0}$ implies that $\mathbf{x}_t = \mathbf{0}$ for all $t \in \mathcal{J}^{(a)}$. %
The proof is complete. %

\subsection{Proof of Theorem \ref{theorem:Cons2-access-repair}}
\label{appendix:proof-Cons2-access-repair}

For the code $\mathcal{C}_1$ constructed by Generic Construction \ref{G-construction1}, every block entry $\mathbf{H}_{i, j}$ of its parity-check matrix $\mathbf{H}$ can be regarded as an $l' \times l'$ block matrix with every block entry an $m \times m$ matrix over $\mathbb{F}_2$. %
For each $0 \leq j \leq k+r-1$, recall that node $\mathbf{c}_{j}$ is denoted as $\mathbf{c}_j = [\mathbf{c}_{j,0}^{\rm T}~\mathbf{c}_{j,1}^{\rm T}~\cdots~\mathbf{c}_{j,l'-1}^{\rm T}]^{\rm T}$, where $\mathbf{c}_{j,i}$ is an $m$-bit column vector. %
For a fixed $i \in \{0, 1, \ldots, r-1\}$, the $i$-th row of the equation $\sum\nolimits_{j = 0}^{k+r-1} \mathbf{H}_{i, j}\mathbf{c}_j = \mathbf{0}$ (defined in \eqref{eqn:def-C}) can be written out. %
We first notice that the equation $\sum\nolimits_{j = 0}^{k+r-1} \mathbf{H}_{i, j}\mathbf{c}_j = \mathbf{0}$ is equivalent to 
\begin{equation}
\label{eqn:opt-access-1}
\sum\nolimits_{v' = 0}^{g-1}\sum\nolimits_{u'=0}^{s-1}\mathbf{H}_{i, v's+u'}\mathbf{c}_{v's+u'} + \sum\nolimits_{u'=0}^{k+r-sg-1}\mathbf{H}_{i, sg+u'}\mathbf{c}_{sg+u'} = \mathbf{0}. %
\end{equation}

Based on \eqref{eqn: G-construction2} and \eqref{eqn: G-construction2-1}, if $a_{v'} \neq u'$, then the $a$-th block row of $\mathbf{H}_{i,v's+u'}$ contains a single nonzero block entry, which is either $\mathbf{A}_{i, v's+u'}\mathbf{\Psi}_1$ or $\mathbf{A}_{i, v's+u'}\mathbf{\Psi}_2$, located in the $a$-th block column. %
Conversely, if $a_{v'} = u'$, the $a$-th block row of $\mathbf{H}_{i,v's+u'}$ contains exactly $s$ nonzero block entries. %
In particular, when $0 \leq w < u'$, $\mathbf{A}_{i, v's+w}\mathbf{\Psi}_3$ only appears in the $a$-th block row and $a(v', w)$-th block column of $\mathbf{H}_{i, v's+u'}$; when $w = u'$, $\mathbf{A}_{i, v's+u'}$ only appears in the $a$-th block row and $a$-th block column of $\mathbf{H}_{i, v's+u'}$; and when $u' < w \leq s-1$, $\mathbf{A}_{i, \sigma(v's+w)}\mathbf{\Psi}_4$ only appears in the $a$-th block row and $a(v', w)$-th block column of $\mathbf{H}_{i, v's+u'}$. %
Recall that $\sigma(v's+w)$ denotes $v's+w \bmod k+r$. %
Thus, the $a$-th block row of \eqref{eqn:opt-access-1} can be written as \eqref{eqn:proof_access-mK}. %
It is clear that the first summation term in both parentheses corresponds to the case $b = a(v',w)$ for $0 \leq w < u'$, the second summation to $a = b$, and the third summation to $b = a(v',w)$ for $u' < w \leq s-1$. %
Moreover, in \eqref{eqn:proof_access-mK}, we have $\mathbf{\Psi} = \mathbf{\Psi}_1$ if $a_{v'} < u'$, $\mathbf{\Psi} = \mathbf{I}_m$ if $a_{v'} = u'$, and $\mathbf{\Psi} = \mathbf{\Psi}_2$ if $a_{v'} > u'$. %
Our aim is to recover the node $\mathbf{c}_{vs+u}$ by the data chunks in $\mathcal{M}^{(v, u)}$. %

\begin{figure*}[t]
\begin{align}
&\scalematrix{\sum\nolimits_{v' = 0}^{g-1} \left( \sum\nolimits_{u'=0}^{a_{v'}-1}\mathbf{A}_{i, v's+u'}\mathbf{\Psi}_3\mathbf{c}_{v's+a_{v'}, a(v',u')}  + \sum\nolimits_{u'=0}^{s-1}\mathbf{A}_{i, v's+u'}\mathbf{\Psi}\mathbf{c}_{v's+u', a} + \sum\nolimits_{u' = a_{v'} +1}^{s-1}\mathbf{A}_{i, v's+u'}\mathbf{\Psi}_4\mathbf{c}_{v's+a_{v'}, a(v',u')} \right) }\nonumber \\
&\scalematrix{ + \left( \sum\nolimits_{u'=0}^{a_{g}-1}\mathbf{A}_{i, gs+u'}\mathbf{\Psi}_3\mathbf{c}_{gs+a_{g}, a(g, u')} +  \sum\nolimits_{u' = 0}^{k+r-sg-1}\mathbf{A}_{i, gs+u'}\mathbf{\Psi}\mathbf{c}_{gs+u', a}  +  
\sum\nolimits_{u' = a_{g} +1}^{s-1}\mathbf{A}_{i, \sigma(gs+u')}\mathbf{\Psi}_4\mathbf{c}_{gs+a_{g}, a(g,u')}
\right) = \mathbf{0}.}   \label{eqn:proof_access-mK} \\[3pt]
&
\begin{bmatrix}
\sum\nolimits_{u'=0}^{u-1}\mathbf{A}_{0, vs+u'}\mathbf{\Psi}_3\mathbf{c}_{vs+u, a(v,u')} + \mathbf{A}_{0, vs+u}\mathbf{c}_{vs+u, a} + \sum\nolimits_{u' = u+1 }^{s-1}\mathbf{A}_{0, \sigma(vs+u')}\mathbf{\Psi}_4\mathbf{c}_{vs+u, a(v,u')}\\
\sum\nolimits_{u'=0}^{u-1}\mathbf{A}_{1, vs+u'}\mathbf{\Psi}_3\mathbf{c}_{vs+u, a(v,u')} + \mathbf{A}_{1, vs+u}\mathbf{c}_{vs+u, a}+ \sum\nolimits_{u' = u+1 }^{s-1}\mathbf{A}_{1, \sigma(vs+u')}\mathbf{\Psi}_4\mathbf{c}_{vs+u, a(v,u')} \\
\vdots \\ \sum\nolimits_{u'=0}^{u-1}\mathbf{A}_{r-1, vs+u'}\mathbf{\Psi}_3\mathbf{c}_{vs+u, a(v,u')} + \mathbf{A}_{r-1, vs+u}\mathbf{c}_{vs+u, a} + \sum\nolimits_{u' = u+1 }^{s-1}\mathbf{A}_{r-1, \sigma(vs+u')}\mathbf{\Psi}_4\mathbf{c}_{vs+u, a(v,u')}
\end{bmatrix} = \nonumber \\
&\begin{bmatrix}
\mathbf{A}_{0,vs}\mathbf{\Psi}_3 & \cdots & \mathbf{A}_{0,vs+u-1}\mathbf{\Psi}_3 & \mathbf{A}_{0,vs+u} & \mathbf{A}_{0,\sigma(vs+u+1)}\mathbf{\Psi}_4 & \cdots & \mathbf{A}_{0,\sigma(vs+s-1)}\mathbf{\Psi}_4 \\
\mathbf{A}_{1,vs}\mathbf{\Psi}_3 & \cdots & \mathbf{A}_{1,vs+u-1}\mathbf{\Psi}_3 & \mathbf{A}_{1,vs+u} & \mathbf{A}_{1,\sigma(vs+u+1)}\mathbf{\Psi}_4 & \cdots & \mathbf{A}_{1,\sigma(vs+s-1)}\mathbf{\Psi}_4 \\
\vdots & \vdots & \vdots  & \vdots & \vdots & \vdots & \vdots \\
\mathbf{A}_{r-1,vs}\mathbf{\Psi}_3 & \cdots & \mathbf{A}_{r-1,vs+u-1}\mathbf{\Psi}_3 & \mathbf{A}_{r-1,vs+u} & \mathbf{A}_{r-1,\sigma(vs+u+1)}\mathbf{\Psi}_4 & \cdots & \mathbf{A}_{r-1,\sigma(vs+s-1)}\mathbf{\Psi}_4 \\
\end{bmatrix} \cdot \nonumber \\
&\begin{bmatrix} \mathbf{c}_{vs+u, a(v, 0)}^{\rm T} ~ \ldots ~ \mathbf{c}_{vs+u, a(v, u-1)}^{\rm T} ~ \mathbf{c}_{vs+u, a}^{\rm T} ~ \mathbf{c}_{vs+u, a(v, u+1)}^{\rm T} ~ \ldots ~ \mathbf{c}_{vs+u, a(v, s-1)}^{\rm T} \end{bmatrix}^{\rm T}. 
\label{eqn:proof_access-mK-parti2}
\end{align}
\end{figure*}

First, we consider the case that $s = r$. %
For every $0 \leq i \leq r-1$ and all $a$ satisfying $a_v = u$, all terms in \eqref{eqn:proof_access-mK}, excluding the $s$ terms on the left-hand side of \eqref{eqn:proof_access-mK-parti2}, can be found from the data chunks in $\mathcal{M}^{(v, u)}$. %
According to \eqref{eqn:proof_access-mK-parti2}, since the $(K+r, K, m)$ base code is MDS, the data chunks $\{ \mathbf{c}_{vs+u, a(v, u')}~:~0 \leq u' \leq s-1\}$ can be determined from $\{ \sum\nolimits_{u'=0}^{u-1}\mathbf{A}_{i, vs+u'}\mathbf{\Psi}_3\mathbf{c}_{vs+u, a(v,u')} + \mathbf{A}_{i, vs+u}\mathbf{c}_{vs+u, a} + \sum\nolimits_{u' = u+1 }^{s-1}\mathbf{A}_{i, \sigma(vs+u')}\mathbf{\Psi}_4\mathbf{c}_{vs+u, a(v,u')}~:~0 \leq i \leq r-1 \}$ for all $0 \leq a \leq l'-1$. %
Specifically, the data chunks
\begin{equation*}
\begin{aligned}
&\{ \mathbf{c}_{vs+u, a}~:~ 0 \leq a \leq l'-1\} =  \{ \mathbf{c}_{vs+u, a(v, u')}~:~a_v = u, 0 \leq u' \leq s-1 \}
\end{aligned}
\end{equation*}
are uniquely determined by the data chunks 
\begin{equation*}
\begin{aligned}
\scalematrix{
\{ \sum\nolimits_{u'=0}^{u-1}\mathbf{A}_{i, vs+u'}\mathbf{\Psi}_3\mathbf{c}_{vs+u, a(v,u')} + \mathbf{A}_{i, vs+u}\mathbf{c}_{vs+u, a} +  \sum\nolimits_{u' = u+1 }^{s-1}\mathbf{A}_{i, \sigma(vs+u')}\mathbf{\Psi}_4\mathbf{c}_{vs+u, a(v,u')}~:~ a_v = u,  ~0 \leq i \leq r-1 \}. }%
\end{aligned}
\end{equation*}
As mentioned above, the values $\{ \sum\nolimits_{u'=0}^{u-1}\mathbf{A}_{i, vs+u'}\mathbf{\Psi}_3\mathbf{c}_{vs+u, a(v,u')} + \mathbf{A}_{i, vs+u}\mathbf{c}_{vs+u, a} + \sum\nolimits_{u' = u+1 }^{s-1}\mathbf{A}_{i, \sigma(vs+u')}\mathbf{\Psi}_4\mathbf{c}_{vs+u, a(v,u')}~:~ a_v = u, ~0 \leq i \leq r-1 \}$ are uniquely determined by the data chunks in $\mathcal{M}^{(v, u)}$. %
Consequently, $\{ \mathbf{c}_{vs+u, a}~:~ 0 \leq a \leq l'-1\}$ are uniquely determined by data chunks in $\mathcal{M}^{(v, u)}$, that is, the node $\mathbf{c}_{vs+u}$ can be recovered from the data chunks in $\mathcal{M}^{(v, u)}$. %

Next, we analyze the case that $s < r$. %
Notice that in this case, $r-s$ nodes do not participate in the repair process of node $\mathbf{c}_{vs+u}$. %
For each $0 \leq j' \leq r-s-1$, the index which corresponds to a non-participating node can be expressed as $j_{j'} = v_{j'}s+u_{j'}$. %
The index set $\mathcal{J}$ is then given by $\mathcal{J} = \{j_{j'} ~:~ 0 \leq j' \leq r-s-1 \}$. %
We futher define an $r \times (r-s+1)$ block sub-matrix $\mathbf{H}_1$ of $\mathbf{H}$ as
\begin{equation*}
\mathbf{H}_1 = 
\begin{bmatrix}
\mathbf{H}_{0, vs+u} & \mathbf{H}_{0, j_0} & \mathbf{H}_{0, j_1} & \cdots & \mathbf{H}_{0, j_{r-s-1}} \\
\mathbf{H}_{1, vs+u} & \mathbf{H}_{1, j_0} & \mathbf{H}_{1, j_1} & \cdots & \mathbf{H}_{1, j_{r-s-1}} \\
\vdots & \vdots & \vdots & \vdots & \vdots \\
\mathbf{H}_{r-1, vs+u} & \mathbf{H}_{r-1, j_0} & \mathbf{H}_{r-1, j_1} & \cdots & \mathbf{H}_{r-1, j_{r-s-1}} \\
\end{bmatrix}.
\end{equation*}
Analogous to the block permutation matrix $\mathbf{P}$ defined in \eqref{eqn:P-MDS}, we define a block permutation matrix $\mathbf{P}' = [\mathbf{P}'_{i,j}]_{0 \leq i, j \leq (r-s+1)l'-1}$ such that
\begin{equation*}
\mathbf{P}'_{i,j} = \mathbf{I}_{m} ~{\rm iff}~i = (r-s+1)j-((r-s+1)l'-1)\left\lfloor \frac{j}{l'}\right\rfloor.
\end{equation*}
Let $\mathbf{H}_2 = \mathbf{P}'\mathbf{H}_1$. %
For each $a \in \{a: a_v = u, 0 \leq a \leq l'-1\}$, define $\mathbf{H}_2^{(a)}$ as the $r \times (r-s+1)l'$ block sub-matrix of $\mathbf{H}_2$ formed by block rows $ar, ar+1, \cdots, ar+r-1$, where every block entry of $\mathbf{H}_2^{(a)}$ is an $m \times m$ matrix over $\mathbb{F}_2$. %
We further define $\mathcal{U}_1^{(a)} \subset \{ 0, 1, \cdots, (g+1)s-1\}$ to be the set of all indices $j$ such that $\mathbf{H}_2^{(a)}$ contains a nonzero block column equal to some block vector of $\{\mathbf{L}_j,\mathbf{L}_j\mathbf{\Psi}_1, \mathbf{L}_j\mathbf{\Psi}_2, \mathbf{L}_j\mathbf{\Psi}_3, \mathbf{L}_j\mathbf{\Psi}_4\}$, where $\mathbf{L}_j$ is defined in \eqref{eqn:L}. %
From \eqref{eqn: G-construction2} and \eqref{eqn: G-construction2-1}, for each $a \in \{a: a_v = u, 0 \leq a \leq l'-1\}$, the $a$-th block row of $\mathbf{H}_{i, vs+u}$ contains $s$ nonzero block entries. %
In particular, when $0 \leq w < u$, $\mathbf{A}_{i, vs+w}\mathbf{\Psi}_3$ only appears in the $a$-th block row and $a(v, w)$-th block column of $\mathbf{H}_{i, vs+u}$; %
and when $w = u$, $\mathbf{A}_{i, vs+u}$ only appears in the $a$-th block row and $a$-th block column of $\mathbf{H}_{i, vs+u}$; %
and when $u < w \leq s-1$, $\mathbf{A}_{i, \sigma(vs+w)}\mathbf{\Psi}_4$ only appears in the $a$-th block row and $a(v, w)$-th block column of $\mathbf{H}_{i, vs+u}$. %
The $a$-th block row of $\mathbf{H}_{i, j_{j'}}$ contains at least one nonzero block entry, which is either $\mathbf{A}_{i,j_{j'}}$, $\mathbf{A}_{i,j_{j'}}\mathbf{\Psi}_1$ or $\mathbf{A}_{i,j_{j'}}\mathbf{\Psi}_2$,  located in the $a$-th block column when $a \in \{a: a_v = u, 0 \leq a \leq l'-1\}$. %
Consequently, for each $a \in \{a: a_v = u, 0 \leq a \leq l'-1\}$, $\{ vs, vs+1, \ldots, vs+s-1\} \cup \mathcal{J}$ is contained in $\mathcal{U}_1^{(a)}$, we can obtain that $|\mathcal{U}_1^{(a)}| \geq r$. %
Using a similar method of the proof of \eqref{eqn:appen-proof-MDS2-00}, one can readily verify that for the given index set $\mathcal{J}$, there exists at least one element $a \in \{a: a_v = u, 0 \leq a \leq l'-1\}$ such that $|\mathcal{U}_1^{(a)}| = r$. %

Now, we will argue by induction on the cardinality of the set $\mathcal{U}_1^{(a)}$ to justify that $\mathbf{c}_{vs+u,a}$ can be obtained based on the data chunks in $\mathcal{M}^{(v, u)}$ for all $0 \leq a \leq l'-1$. %
Specifically, the data chunks 
\begin{equation*}
\begin{aligned}
&\{ \mathbf{c}_{vs+u, a}~:~ 0 \leq a \leq l'-1\} =  \{ \mathbf{c}_{vs+u, a(v, u')}~:~a_v = u,~0 \leq u' \leq s-1 \}.
\end{aligned}
\end{equation*}
To establish the induction basis, let $a \in \{a: a_v = u, 0 \leq a \leq l'-1\}$ be one of the values that satisfy $|\mathcal{U}_1^{(a)}| = r$, 
we can obtain that
\begin{equation}
\label{eqn:M-optimal-access}
\small
\begin{aligned} 
&\scalematrix{\begin{bmatrix}
\mathbf{A}_{0,vs}\mathbf{\Psi}_3 & \cdots & \mathbf{A}_{0,vs+u-1}\mathbf{\Psi}_3 & \mathbf{A}_{0,vs+u} & \mathbf{A}_{0,\sigma(vs+u+1)}\mathbf{\Psi}_4 & \cdots & \mathbf{A}_{0,\sigma(vs+u+s-1)}\mathbf{\Psi}_4 & \mathbf{A}_{0, j_0} & \cdots & \mathbf{A}_{0, j_{r-s-1}}\\
\mathbf{A}_{1,vs}\mathbf{\Psi}_3 & \cdots & \mathbf{A}_{1,vs+u-1}\mathbf{\Psi}_3 & \mathbf{A}_{1,vs+u} & \mathbf{A}_{1,\sigma(vs+u+1)}\mathbf{\Psi}_4 & \cdots & \mathbf{A}_{1,\sigma(vs+u+s-1)}\mathbf{\Psi}_4 & \mathbf{A}_{1, j_0} & \cdots & \mathbf{A}_{1, j_{r-s-1}}\\
\vdots & \vdots & \vdots  & \vdots & \vdots & \vdots & \vdots & \vdots & \vdots & \vdots \\
\mathbf{A}_{r-1,vs}\mathbf{\Psi}_3 & \cdots & \mathbf{A}_{r-1,vs+u-1}\mathbf{\Psi}_3 & \mathbf{A}_{r-1,vs+u} & \mathbf{A}_{r-1,\sigma(vs+u+1)}\mathbf{\Psi}_4 & \cdots & \mathbf{A}_{r-1,\sigma(vs+u+s-1)}\mathbf{\Psi}_4 & \mathbf{A}_{r-1, j_0} & \cdots & \mathbf{A}_{r-1, j_{r-s-1}}\\
\end{bmatrix} }\cdot \\
&\begin{bmatrix} \mathbf{c}_{vs+u, a(v, 0)}^{\rm T} & \cdots & \mathbf{c}_{vs+u, a(v, u-1)}^{\rm T} & \mathbf{c}_{vs+u, a}^{\rm T} & \mathbf{c}_{vs+u, a(v, u+1)}^{\rm T} & \cdots & \mathbf{c}_{vs+u, a(v, s-1)}^{\rm T} & \mathbf{b}_0^{\rm T} & \cdots & \mathbf{b}_{r-s-1}^{\rm T} \end{bmatrix}^{\rm T} \\
& = \begin{bmatrix}
\mathbf{C}_0^{\rm T} & \mathbf{C}_1^{\rm T} & \cdots & \mathbf{C}_{r-1}^{\rm T}
\end{bmatrix}^{\rm T}. %
\end{aligned}
\end{equation} 
Here, for each $0 \leq i \leq r-1$, $\mathbf{C}_i = \sum\nolimits_{u'=0}^{u-1}\mathbf{A}_{i, vs+u'}\mathbf{\Psi}_3\mathbf{c}_{vs+u, a(v,u')} + \mathbf{A}_{i, vs+u}\mathbf{c}_{vs+u, a} + \sum\nolimits_{u' = u+1 }^{s-1}\mathbf{A}_{i, \sigma(vs+u')}\mathbf{\Psi}_4\mathbf{c}_{vs+u, a(v,u')} + \sum\nolimits_{j' = 0}^{r-s-1}\mathbf{A}_{i, j_{j'}}\mathbf{b}_{j'}$ is determined by the data chunks in $\mathcal{M}^{(v, u)}$. %
One can easily verify that $\{ 0, 1, \cdots, r-s-1\}$ can be partitioned into three disjoint subsets, $\mathcal{W}_1$, $\mathcal{W}_2$ and $\mathcal{W}_3$, in a manner similar to the partition in \eqref{eqn:W-pert}. %
Thus, if $j' \in \mathcal{W}_1$, then $\mathbf{b}_{j'} = \mathbf{c}_{v_{j'}s+u_{j'}, a}$; if $j' \in \mathcal{W}_2$, then $\mathbf{b}_{j'} = \mathbf{\Psi}_2\mathbf{c}_{v_{j'}s+u_{j'}, a}+\mathbf{\Psi}_3\mathbf{c}_{v_{p}s+u_{p}, a(v_{j'}, u_{j'})}$; if $j' \in \mathcal{W}_3$, then $\mathbf{b}_{j'} = \mathbf{\Psi}_1\mathbf{c}_{v_{j'}s+u_{j'}, a}+\mathbf{\Psi}_4\mathbf{c}_{v_{p}s+u_{p}, a(v_{j'}, u_{j'})}$. %
Since the first matrix on the left-hand side of \eqref{eqn:M-optimal-access} is a full-rank matrix, for each $a \in \{a: a_v = u, 0 \leq a \leq l'-1\}$ such that $|\mathcal{U}_1^{(a)}| = r$, $\{ \mathbf{c}_{vs+u, a(v,u')}~:~0 \leq u' \leq s-1 \}$ can be obtained based on the data chunks in $\mathcal{M}^{(v, u)}$. %

In what follows, assume that for all $a \in \{a: a_v = u, 0 \leq a \leq l'-1\}$ with $|\mathcal{U}_1^{(a)}| \leq w-1$, the data chunks $\{ \mathbf{c}_{v_{j'}s+u_{j'}, a}~:~0 \leq j' \leq r-s-1 \}$ have been obtained. %
We now proceed to show that for all $a \in \{a: a_v = u, 0 \leq a \leq l'-1\}$ with $|\mathcal{U}_1^{(a)}| = w$, the data chunks $\{ \mathbf{c}_{vs+u, a(v,u')}~:~0 \leq u' \leq s-1 \}$ is also determined by the data chunks in $\mathcal{M}^{(v, u)}$. %
Let $a$ be an element in $\{a: a_v = u, 0 \leq a \leq l'-1\}$ such that $|\mathcal{U}_1^{(a)}| = w$. %
For any $j \in \mathcal{U}_1^{(a)} \setminus (\{ vs, vs+1, \cdots, vs+s-1\} \cup \mathcal{J})$, there exists a unique $j' \in \{0, 1, \cdots, r-s-1 \}$ such that $a_{v_{j'}} = u_{j'}$ and $0 \leq j-v_{j'}s \leq s-1$. %
Now consider the matrix $\mathbf{H}_2^{(a(v_{j'},j-v_{j'}s))}$. %
It can be observed that for the chosen $j$, no $p \in \{0, 1, \ldots, r-s-1\}$ simultaneously satisfies $v_p = v_{j'}$ and $u_p = j-v_{j'}s$. %
Consequently, we have $\mathcal{U}_1^{(a(v_{j'}, j-v_{j'}s))} \subset \mathcal{U}_1^{(a)}$ and $j \notin \mathcal{U}_1^{(a(v_{j'}, j-v_{j'}s))}$, which implies  $|\mathcal{U}_1^{(a(v_{j'}, j-v_{j'}s))}| \leq w-1$. %
We can conclude that the induction hypothesis applies and $\mathbf{c}_{v_{j'}s+u_{j'}, a(v_{j'},j-v_{j'}s)}$ has been obtained for the given $a$. %
Rephrasing this, we can verify that for every $j \in \mathcal{U}_1^{(a)} \setminus (\{ vs, vs+1, \cdots, vs+s-1\} \cup \mathcal{J})$, the data chunk $\mathbf{c}_{v_{j'}s+u_{j'}, a(v_{j'}, j-v_{j'}s)}$ has been obtained. %
Therefore, based on \eqref{eqn:M-optimal-access} again, for the given $a$, the data chunks $\{ \mathbf{c}_{vs+u, a(v,u')}~:~0 \leq u' \leq s-1 \}$ can be determined by the data chunks in $\mathcal{M}^{(v, u)}$. %
We thus establish that for all $a \in \{a: a_v = u, 0 \leq a \leq l'-1\}$ with $|\mathcal{U}_1^{(a)}|  = w$, the data chunks $\{ \mathbf{c}_{vs+u, a(v,u')}~:~0 \leq u' \leq s-1 \}$ is determined by the data chunks in $\mathcal{M}^{(v, u)}$. %
As a result, node $\mathbf{c}_{vs+u}$ can also be recovered from the data chunks in $\mathcal{M}^{(v, u)}$ when $s < r$. %
We conclude that, for the code $\mathcal{C}_1$ constructed by Generic Construction \ref{G-construction1}, node $\mathbf{c}_{vs+u}$ can be recovered from the data chunks in $\mathcal{M}^{(v, u)}$. %
The proof is complete. %

\subsection{Proof of Theorem \ref{theorem:MDS-2}}
\label{appendix:proof-MDS-2}
We begin by selecting a $(K+r, K, m)$ binary MDS array code where $r > 2$ is even as the base code. %
Recall that for defined $k$ and $r$, $s = \frac{r}{2}$, $g = \frac{k+r}{s+1}$ and $l = ms^g$. Let $l' =s^g$. %
Consider arbitrary $j_0, j_1, \ldots, j_{r-1}$ subject to $0 \leq j_0 < j_1 < \ldots < j_{r-1} \leq k+r-1$, and let $\mathbf{H}'$ denote the $r \times r$ block sub-matrix of $\mathbf{H}$ as defined in \eqref{eqn:submatrix-H}, that is, $\mathbf{H}'$ is obtained from $\mathbf{H}$ by restricting to the block columns indexed by $j_0, j_1, \ldots, j_{r-1}$. %
In order to prove the MDS property of $\mathcal{C}_2$, according to proposition \ref{prop:pre-MDS}, it is equivalent to prove that $\mathbf{H}'$, when regarded as an $rl \times rl$ matrix over $\mathbb{F}_2$, is full rank. %
To establish full rank of $\mathbf{H}'$, we show that for an $rl$-bit column vector $\mathbf{x}$, 
\begin{equation}
\label{eqn:H'x = 0_appendix}
\mathbf{H}'\mathbf{x} = \mathbf{0}~{\rm implies}~\mathbf{x} = \mathbf{0}.
\end{equation}
Here, $\mathbf{x}$ is expressed as $\mathbf{x} = [\mathbf{x}_0^{\rm T}~\mathbf{x}_1^{\rm T}~\ldots~\mathbf{x}_{rl'-1}^{\rm T}]^{\rm T}$, where each coordinate $\mathbf{x}_i$ is an $m$-bit column vector. %
Let $\mathbf{H}'' = \mathbf{P}\mathbf{H}'$ where $\mathbf{P}$ is defined in \eqref{eqn:P-MDS}. %
In order to prove \eqref{eqn:H'x = 0_appendix}, it is equivalent to show that
\begin{equation}
\label{eqn:H''x = 0_appendix}
\mathbf{H}''\mathbf{x} = \mathbf{0} ~{\rm implies} ~\mathbf{x} = \mathbf{0}. %
\end{equation}

For each $a \in \{0, 1, \ldots, l'-1 \}$, define $\mathbf{H}''^{(a)}$ as the $r \times rl'$ block sub-matrix of $\mathbf{H}''$ formed by block rows $ar, ar+1, \ldots, ar+r-1$, where each block entry is an $m \times m$ matrix over $\mathbb{F}_2$. %
Further, for each $0 \leq j \leq 2sg-1$, define a block column vector $\mathbf{L}_{j} = \left[ \mathbf{A}_{0,j}^{\rm T}~\mathbf{A}_{1,j}^{\rm T}~\ldots~\mathbf{A}_{r-1,j}^{\rm T} \right]^{\rm T}$. %
Analogous to the sets $\mathcal{U}^{(a)}$, $\mathcal{J}^{(a)}$, and $\mathcal{J}^{(a)}(j)$ defined in the proof of MDS property for $\mathcal{C}_1$, we now define subsets of block column indices for the block matrix $\mathbf{H}''^{(a)}$ for every $0 \leq a \leq l'-1$:
\begin{itemize}
  \item $\hat{\mathcal{U}}^{(a)} = \{0 \leq j \leq 2sg-1~:~\mathbf{H}''^{(a)} ~{\rm contains~a~nonzero~block~column~belonging~to}~\{ \mathbf{L}_{j},\mathbf{L}_{j}\mathbf{\Psi}_1,\mathbf{L}_{j}\mathbf{\Psi}_2,\mathbf{L}_{j}\mathbf{\Psi}_3,\\ \mathbf{L}_{j}\mathbf{\Psi}_4\} \}$; 
  \item $\hat{\mathcal{J}}^{(a)} \subset \{ 0, 1, \ldots, rl'-1\}$ lists the indices of nonzero block columns in $\mathbf{H}''^{(a)}$;
  \item For each $0 \leq j \leq 2sg-1$, $\hat{\mathcal{J}}^{(a)}(j) = \{t \in \hat{\mathcal{J}}^{(a)}~:~t\textrm{-}{\rm th}~ {\rm block~column ~of~} \mathbf{H}''^{(a)} {\rm ~belongs~to~} \{ \mathbf{L}_{j},\mathbf{L}_{j}\mathbf{\Psi}_1,\mathbf{L}_{j}\mathbf{\Psi}_2,\\ \mathbf{L}_{j}\mathbf{\Psi}_3,\mathbf{L}_{j}\mathbf{\Psi}_4\}\}$. %
\end{itemize}

In order to prove \eqref{eqn:H''x = 0_appendix}, it is equivalent to prove that for each $0 \leq a \leq l'-1$, 
\begin{equation}
\label{eqn:H''ax=0}
\mathbf{H}''^{(a)}[\mathbf{x}_0^{\rm T}~\mathbf{x}_1^{\rm T}~\ldots~\mathbf{x}_{rl'-1}^{\rm T}]^{\rm T} = \mathbf{0} ~ {\rm implies~that~}  \mathbf{x}_t = \mathbf{0}~{\rm for~all}~t \in \hat{\mathcal{J}}^{(a)}.
\end{equation}
In what follows, we will proceed by induction on the cardinality of the set $\hat{\mathcal{U}}^{(a)}$ to prove that \eqref{eqn:H''ax=0} holds for each $0 \leq a \leq l'-1$. %

First, using a very similar approach as proving \eqref{eqn:appen-proof-MDS2-00}, we can prove that for any $0 \leq j_0 < j_1 < \ldots < j_{r-1} \leq k+r-1$, 
\begin{equation*}
\min\limits_{ a \in \{0, 1, \ldots, l'-1 \} } |\hat{\mathcal{U}}^{(a)}| = r.
\end{equation*}
Next, To establish the induction basis we need to prove that for all $a$ such that $|\hat{\mathcal{U}}^{(a)}| = r$, \eqref{eqn:H''ax=0} holds.
Let $a$ be one of the values that satisfy $|\hat{\mathcal{U}}^{(a)}| = r$. %
Let us sort the $r$ number in $\hat{\mathcal{U}}^{(a)}$ in ascending order, and denote each number as $i_h$ with $0 \leq h \leq r-1$. %
For the given $a$, all the nonzero block columns of $\mathbf{H}''^{(a)}$ belong to the set
\begin{equation*}
\{ \mathbf{L}_{i_0}\mathbf{\Psi}, \mathbf{L}_{i_1}\mathbf{\Psi}, \ldots, \mathbf{L}_{i_{r-1}}\mathbf{\Psi}: \mathbf{\Psi} \in \{\mathbf{I}_m, \mathbf{\Psi}_1, \mathbf{\Psi}_2, \mathbf{\Psi}_3, \mathbf{\Psi}_4 \} \}.
\end{equation*}
Thus, $\mathbf{H}''^{(a)}[\mathbf{x}_0^{\rm T}~\mathbf{x}_1^{\rm T}~\ldots~\mathbf{x}_{rl'-1}^{\rm T}]^{\rm T} = \mathbf{0}$ implies  
\begin{equation*}
[\mathbf{L}_{i_0}~\mathbf{L}_{i_1}~\ldots~\mathbf{L}_{i_{r-1}}] [\bar{\mathbf{x}}_{i_0}^{\rm T}~\bar{\mathbf{x}}_{i_1}^{\rm T}~\ldots~\bar{\mathbf{x}}_{i_{r-1}}^{\rm T}]^{\rm T} = \mathbf{0}. %
\end{equation*}
Since $\bigcup_{0 \leq h  \leq r-1} \hat{\mathcal{J}}^{(a)}(i_h) = \hat{\mathcal{J}}^{(a)}$, where each $\hat{\mathcal{J}}^{(a)}(i_h)$ contains either one or two indices. %
For $0 \leq h\leq r-1$ 
\begin{itemize}
\item if $|\hat{\mathcal{J}}^{(a)}(i_h)| = 1$ with $\hat{\mathcal{J}}^{(a)}(i_h) = \{ t \}$, then $\bar{\mathbf{x}}_{i_h} = \mathbf{x}_t$;
\item if $|\hat{\mathcal{J}}^{(a)}(i_h)| = 2$ with $\hat{\mathcal{J}}^{(a)}(i_h) = \{t, t'\}$, then $\bar{\mathbf{x}}_{i_h}$ is a linear combination of $\mathbf{x}_t$ and $\mathbf{x}_{t'}$. %
\end{itemize}
Since the block columns $\mathbf{L}_{i_0}, \mathbf{L}_{i_1}, \ldots, \mathbf{L}_{i_{r-1}}$ form the block sub-matrix consisting of block columns $i_0, i_1, \ldots, i_{r-1}$ of the parity-check matrix $\mathbf{A}$ of the $(K+r, K, m)$ base code, the MDS property of the base code implies that $\bar{\mathbf{x}}_{i_h} = \mathbf{0}$ for all $0 \leq h \leq r-1$. %
In the following, we determine the specific indices $t$ involved in each case and show that $\mathbf{x}_t = \mathbf{0}$ for all $t \in \hat{\mathcal{J}}^{(a)}$.

Based on \eqref{eqn: G-cons-optrepair} and \eqref{eqn:G-cons-optrepair-1}, when $0 \leq u \leq s-1$, all nonzero block entries of $\mathbf{H}_{i,j}$ belong to the set $\scalematrix{ \{ \mathbf{A}_{i, 2j-2v-2u}\mathbf{\Psi}_3, \ldots, \mathbf{A}_{i, 2j-2v-u-1}\mathbf{\Psi}_3,  \mathbf{A}_{i, 2j-2v-u}\mathbf{\Psi}_1, \mathbf{A}_{i, 2j-2v-u}, \mathbf{A}_{i, 2j-2v-u}\mathbf{\Psi}_2, \mathbf{A}_{i, 2j-2v-u+1}\mathbf{\Psi}_4, \mathbf{A}_{i, 2j-2v-u+2}\mathbf{\Psi}_4, \ldots , }\\  \mathbf{A}_{i, 2(j-u-v)+s-1}\mathbf{\Psi}_4 \}$; when $u = s$, all nonzero block entries of $\mathbf{H}_{i, j}$ belong to the set $\{ \mathbf{A}_{i, 2(j-v)-s}, \mathbf{A}_{i, 2(j-v)-s+1}, \\  \ldots, \mathbf{A}_{2(j-v)-1} \}$. %
Moreover, when $0 \leq u \leq s-1$, if $a_v \neq u$, then the $a$-th block row of $\mathbf{H}_{i, j}$ contains only a single nonzero block entry, which is either $\mathbf{A}_{i, 2j-2v-u}\mathbf{\Psi}_1$ or $\mathbf{A}_{i, 2j-2v-u}\mathbf{\Psi}_2$,  and if $a_v = u$, then the $a$-th block row of $\mathbf{H}_{i, j}$ contains $s$ nonzero block entries. %
In particular, when $0 \leq w < u$, $\mathbf{A}_{i, 2(j-u-v)+w}$ only appears in the $a$-th block row and $a(v, w)$-th block column of $\mathbf{H}_{i, j}$; %
and when $w = u$, $\mathbf{A}_{i, 2j-2v-u}$ only appears in the $a$-th block row and $a$-th block column of $\mathbf{H}_{i, j}$; %
and when $u < w \leq s-1$, $\mathbf{A}_{i, 2(j-u-v)+w}\mathbf{\Psi}_4$ only appears in the $a$-th block row and $a(v, w)$-th block column of $\mathbf{H}_{i, j}$. %

According to the above characteristic of $\mathbf{H}_{i, j}$, we define $\hat{\mathcal{W}}$, $\hat{\mathcal{W}}_1$, $\hat{\mathcal{W}}_2$ and $\hat{\mathcal{W}}_3$ as follows. %
Define a set $\hat{\mathcal{W}}$ as $\hat{\mathcal{W}} = \{h: 0 \leq h \leq r-1, i_h \in \{ 2sv+s+t: 0 \leq t \leq s-1, 0 \leq v \leq g-1 \} \}$. %
For every $h \in \{ 0, 1, \ldots, r-1 \} \setminus \hat{\mathcal{W}}$, we write the index $i_h$ in the form
\begin{equation*}
i_h = 2v_hs+u_h,
\end{equation*}
where $0 \leq v_h \leq g-1$ and $0 \leq u_h \leq s-1$. %
Let us further partition $\{ 0, 1, \ldots, r-1 \} \setminus \hat{\mathcal{W}}$ into three disjoint subsets $\hat{\mathcal{W}}_1$, $\hat{\mathcal{W}}_2$ and $\hat{\mathcal{W}}_3$, which are defined in a same way as $\mathcal{W}_1$, $\mathcal{W}_2$ and $\mathcal{W}_3$ (defined in \eqref{eqn:W-pert}). %
Redefine $\hat{\mathcal{W}}_1$ by including the elements of $\hat{\mathcal{W}}$ as $\hat{\mathcal{W}}_1 = \hat{\mathcal{W}}_1 \cup \hat{\mathcal{W}}$. %

It is clear that for each $h \in \mathcal{W}_1$, we have $\mathcal{J}^{(a)}(i_h) = \{ hl'+a\}$. %
Thus, we conclude that 
\begin{equation*}
\bar{\mathbf{x}}_{i_h} = \mathbf{x}_{hl'+a} = \mathbf{0}
\end{equation*}
for all $h \in \mathcal{W}_1$. %

Considering the case $h \in \mathcal{W}_2$, observe that $\mathcal{J}^{(a)}(i_h) = \{ hl'+a, pl'+a(v_h, u_h)\}$ which implies that 
\begin{equation}
\label{eqn:append-proof-MDS2-W2-1}
\bar{\mathbf{x}}_{i_h} = \mathbf{\Psi}_2\mathbf{x}_{hl'+a} + \mathbf{\Psi}_3\mathbf{x}_{pl'+a(v_h, u_h)} = \mathbf{0}.
\end{equation}
Let us consider $\mathbf{H}''^{(a(v_h, u_h))}$, whose nonzero block columns are described as follows:
\begin{equation*}
|\mathcal{U}^{(a(v_h, u_h))}| = r,
\end{equation*}
\begin{equation*}
\mathcal{J}^{(a(v_h, u_h))}(i_p) = \{ hl'+a, pl'+a(v_h, u_h) \}.
\end{equation*}
From this we obtain that
\begin{equation}
\label{eqn:append-proof-MDS2-W2-2}
\mathbf{\Psi}_4\mathbf{x}_{hl'+a} + \mathbf{\Psi}_1\mathbf{x}_{pl'+a(v_h, u_h)} = \mathbf{0}.
\end{equation}
Recall that $\left[ \begin{smallmatrix} \mathbf{\Psi}_1 & \mathbf{\Psi}_4 \\ \mathbf{\Psi}_3 & \mathbf{\Psi}_2 \end{smallmatrix} \right]$ has full rank $2m$, \eqref{eqn:append-proof-MDS2-W2-1} and \eqref{eqn:append-proof-MDS2-W2-2} imply that $\mathbf{x}_{hl'+a} = \mathbf{x}_{pl'+a(v_h, u_h)} = \mathbf{0}$. %
Using a very similar analysis for $\mathcal{W}_3$ to the one used for $\mathcal{W}_2$, we conclude that for each $h \in \mathcal{W}_3$, $\mathbf{x}_{hl'+a} = \mathbf{x}_{pl'+a(v_h, u_h)} = \mathbf{0}$. %
This completes the induction basis, proving that $\mathbf{x}_t = \mathbf{0}$ for all $t \in \hat{\mathcal{J}}^{(a)}$ when $|\hat{\mathcal{U}}^{(a)}| = r$. %

In what follows, assume $\mathbf{x}_t = \mathbf{0}$ for all $t \in \hat{\mathcal{J}}^{(a)}$ and all $a$ with $|\hat{\mathcal{U}}^{(a)}| \leq w-1$, where $w \geq r$. %
We will prove that for all $a$ such that $|\hat{\mathcal{U}}^{(a)}| = w$, $\mathbf{x}_t = \mathbf{0}$ for all $t \in \hat{\mathcal{J}}^{(a)}$. %
This is equivalent to proving that for every $j \in \hat{\mathcal{U}}^{(a)}$, $\mathbf{x}_t = \mathbf{0}$ for all $t \in \hat{\mathcal{J}}^{(a)}(j)$. %
Recall that $\mathbf{H}'$ is obtained from $\mathbf{H}$ by restricting to the block columns indexed by $j_0, j_1, \ldots, j_{r-1}$, and each $j_i$ can be expressed as $j_i = v_i(s+1)+u_i$ with $0 \leq v_i \leq g-1$ and $0 \leq u_i \leq s$. %
When $0 \leq u_i \leq s-1$, we define $j'_i = 2j_i - 2v_i-u_i$, whereas for $u_i = s$, the index $j'_i$ is chosen from the set $\{2j_i - 2v_i-s, 2j_i - 2v_i-s+1, \ldots, 2j_i - 2v_i-1 \}$ according to the value of $a$. %
For any $j \in \hat{\mathcal{U}}^{(a)} \backslash \{ j'_0, j'_1, \ldots, j'_{r-1} \}$, there exists a unique $h \in \{0, 1, \ldots, r-1 \}$ such that $a_{v_h} = u_h$ and $0 \leq j-2v_hs \leq s-1$, with the corresponding set $\hat{\mathcal{J}}^{(a)}(j) = \{ hl'+a(v_h, \alpha)\}$, where $\alpha = j-2v_hs$. %
Consider the matrix $\mathbf{H}''^{(a(v_h, \alpha))}$. %
We observe that for the chosen $j$, no $p \in \{0, 1, \ldots, r-1 \} \backslash \{i: 0 \leq i \leq r-1, j_i \in \{v_i(s+1)+s~:~0 \leq v_i \leq g-1 \}\}$ simultaneously satisfies $v_p = v_h$ and $u_p = \alpha$. %
Thus, we have $\hat{\mathcal{U}}^{(a(v_h, \alpha))} \subset \hat{\mathcal{U}}^{(a)}$ and $j \notin \hat{\mathcal{U}}^{(a(v_h, \alpha))}$, which implies $|\hat{\mathcal{U}}^{(a(v_h, \alpha))}| \leq w-1$. %
We can conclude that the induction hypothesis applies and $\mathbf{x}_t = \mathbf{0}$ for all $t \in \hat{\mathcal{J}}^{(a(v_h, \alpha))}$. %
Furthermore, $\mathbf{x}_{hl'+a(v_h, \alpha)} = \mathbf{0}$ since $hl'+a(v_h, \alpha)$ belongs to  $\hat{\mathcal{J}}^{(a(v_h, \alpha))}$. %
Rephrasing this, we have established that for every $j \in \hat{\mathcal{U}}^{(a)} \backslash \{ j'_0, \ldots, j'_{r-1} \}$, $\mathbf{x}_t = \mathbf{0}$ for all $t \in \hat{\mathcal{J}}^{(a)}(j)$. %

Finally, we consider the variables $\mathbf{x}_t$ for all $t \in \hat{\mathcal{J}}^{(a)}(j'_h)$, where $0 \leq h \leq r-1$. %
Note that $\mathbf{H}''^{(a)}$ reduces to containing only the nonzero columns from the set $\{ \mathbf{L}_{j'_0}\mathbf{\Psi}, \mathbf{L}_{j'_1}\mathbf{\Psi}, \ldots, \mathbf{L}_{j'_{r-1}}\mathbf{\Psi}~:~\mathbf{\Psi} = \mathbf{I}_m, \mathbf{\Psi}_1, \mathbf{\Psi}_2, \mathbf{\Psi}_3, \mathbf{\Psi}_4 \}$. %
We directly conclude that for every $0 \leq h \leq r-1$, $\mathbf{x}_t = \mathbf{0}$ for all $t \in \mathcal{J}^{(a)}(j'_h)$ due to the MDS property of the $(K+r, K, m)$ base code. %
Since the proof of this claim follows exactly the same reasoning as the induction basis proof above, we omit here. %
Consequently, for every $0 \leq a \leq l'-1$, we have $\mathbf{x}_t = \mathbf{0}$ for all $t \in \mathcal{J}^{(a)}$ due to the MDS property of the $(K+r, K, m)$ base code. %
This immediately implies that \eqref{eqn:H''x = 0_appendix} holds, and thus under the base code's MDS assumption, code $\mathcal{C}_2$ is MDS. %
The proof is complete. %

\subsection{Proof of Theorem \ref{theorem:optrepair-M}}
\label{appendix:proof-optrepair-M}
For the code $\mathcal{C}_2$ constructed by Generic Construction \ref{cons:G-cons-optrepair}, every block entry $\mathbf{H}_{i, j}$ of its parity-check matrix $\mathbf{H}$ can be regarded as an $l' \times l'$ block matrix with every block entry an $m \times m$ matrix over $\mathbb{F}_2$. %
For each $0 \leq j \leq k+r-1$, recall that the node $\mathbf{c}_j$ is denoted as $\mathbf{c}_j = [\mathbf{c}_{j,0}^{\rm T}~\mathbf{c}_{j,1}^{\rm T}~\ldots\mathbf{c}_{j,l'-1}^{\rm T}]^{\rm T}$, where each data chunk $\mathbf{c}_{j, i}$ is an $m$-bit column vector. %
For a fixed $i \in \{0, 1, \ldots, r-1\}$, the $i$-th block row of the parity-check equation $\sum\nolimits_{j = 0}^{k+r-1}\mathbf{H}_{i, j}\mathbf{c}_j = \mathbf{0}$ can be written as 
\begin{equation}
\label{eqn:opt-repair-Hc=0}
\sum\nolimits_{v' = 0}^{g-1}\sum\nolimits_{u'=0}^{s}\mathbf{H}_{i, v'(s+1)+u'}\mathbf{c}_{v'(s+1)+u'} = \mathbf{0}. %
\end{equation}

Based on \eqref{eqn: G-cons-optrepair} and \eqref{eqn:G-cons-optrepair-1}, for $0 \leq u' \leq s-1$, if $a_{v'} \neq u'$, the $a$-th block row of $\mathbf{H}_{i, v'(s+1)+u'}$ contains only a single nonzero block entry, which is either $\mathbf{A}_{i, 2v's+u'}\mathbf{\Psi}_1$ or $\mathbf{A}_{i, 2v's+u'}\mathbf{\Psi}_2$; if $a_{v'} = u'$, the $a$-th block row of $\mathbf{H}_{i, v'(s+1)+u'}$ contains $s$ nonzero block entries. %
In particular, when $0 \leq w < u'$, $\mathbf{A}_{i, 2v's+w}$ only appears in the $a$-th block row and $a(v', w)$-th block column of $\mathbf{H}_{i, v'(s+1)+u'}$; %
and when $w = u'$, $\mathbf{A}_{i, 2v's+u'}$ only appears in the $a$-th block row and $a$-th block column of $\mathbf{H}_{i, v'(s+1)+u'}$; %
and when $u' < w \leq s-1$, $\mathbf{A}_{i, 2v's+w}\mathbf{\Psi}_4$ only appears in the $a$-th block row and $a(v', w)$-th block column of $\mathbf{H}_{i, v'(s+1)+u'}$. %
For $u' = s$, the $a$-th block row of $\mathbf{H}_{i, v'(s+1)+u'}$ contains only a single nonzero block entry $\mathbf{A}_{i, 2v's+s+a_{v'}}$. %
Thus, for each $0 \leq a \leq l'-1$, the $a$-th block row of \eqref{eqn:opt-repair-Hc=0} can be written as 
\begin{equation}
\label{eqn:opt-repair-Hc=0-1}
\begin{aligned}
&\sum\nolimits_{v' = 0}^{g-1}\left(\sum\nolimits_{u'=0}^{a_{v'}-1}\mathbf{A}_{i,2v's+u'}\mathbf{\Psi}_3\mathbf{c}_{v'(s+1)+a_{v'},a(v',u')} + \sum\nolimits_{u'=0}^{s-1}\mathbf{A}_{i,2v's+u'}\mathbf{\Psi}\mathbf{c}_{v'(s+1)+u',a} + \right. \\
&\qquad \qquad \qquad \qquad \qquad \qquad \left. \sum\nolimits_{u'=a_{v'}+1}^{s-1}\mathbf{A}_{i, 2v's+u'}\mathbf{\Psi}_4\mathbf{c}_{v'(s+1)+a_{v'},a(v',u')} + \mathbf{A}_{i, 2v's+s+a_{v'}}\mathbf{c}_{v'(s+1)+s,a} \right) = \mathbf{0}, 
\end{aligned}
\end{equation}
where $\mathbf{\Psi} = \mathbf{\Psi}_1$ if $a_{v'} < u'$, $\mathbf{\Psi} = \mathbf{I}_m$ if $a_{v'} = u'$, and $\mathbf{\Psi} = \mathbf{\Psi}_2$ if $a_{v'} > u'$. %

Since the proof for the case $0 \leq u \leq s-1$ is almost the same as that of Theorem \ref{theorem:Cons2-access-repair}, we omit it. %
In what follows, we prove that for $u = s$, node $\mathbf{c}_{v(s+1)+s}$ can be recovered from the data chunks in the set $\mathcal{M}^{(v, s)}$, where $0 \leq v \leq g-1$. %
The $l'$ equations involved in \eqref{eqn:opt-repair-Hc=0} are divided into $\bar{l} = \frac{l'}{s}$ groups, each of $s$ equations. %
To repair node $\mathbf{c}_{v(s+1)+s}$, we sum the block rows (defined in \eqref{eqn:opt-repair-Hc=0-1}) indexed by $a(v,0), a(v,1), \ldots, a(v,s-1)$, obtaining

\begin{equation}
\label{eqn:opt-repair-Hc=0-2}
\begin{aligned}
&\sum\nolimits_{v' \neq v, v' = 0}^{g-1}\left(\sum\nolimits_{u'=0}^{a_{v'}-1}\mathbf{A}_{i,2v's+u'}\mathbf{\Psi}_3\sum\nolimits_{w=0}^{s-1}\mathbf{c}_{v'(s+1)+a_{v'},a(v',u')+ws^{v-1}} + \right. \\
&\left.\sum\nolimits_{u'=0}^{s-1}\mathbf{A}_{i,2v's+u'}\mathbf{\Psi}\sum\nolimits_{w=0}^{s-1}\mathbf{c}_{v'(s+1)+u',a+ws^{v-1}} +\sum\nolimits_{u'=a_{v'}+1}^{s-1}\mathbf{A}_{i, 2v's+u'}\mathbf{\Psi}_4\sum\nolimits_{w=0}^{s-1}\mathbf{c}_{v'(s+1)+a_{v'},a(v',u')+ws^{v-1}}\right)  \\
&\underline{+\sum\nolimits_{u'=0}^{s-1}\mathbf{A}_{i,2vs+u'}\left(\mathbf{\Psi}_1\sum\nolimits_{w=0}^{u'-1}\mathbf{c}_{v(s+1)+u',w}+ 
 \mathbf{c}_{v(s+1)+u',u'}+\mathbf{\Psi}_2\sum\nolimits_{w=1}^{s-1-u'}\mathbf{c}_{v(s+1)+u',u'+w}+ \right. }  \\
&\underline{ \left. \mathbf{\Psi}_4\sum\nolimits_{w=1}^{u'}\mathbf{c}_{v(s+1)+u'-w,u'}+ \mathbf{\Psi}_3\sum\nolimits_{w=1}^{s-1-u'}\mathbf{c}_{v(s+1)+u'+w,u'} \right) + 
\sum\nolimits_{u' = 0}^{s-1}\mathbf{A}_{i,2(v+1)s-s+u'}\mathbf{c}_{v(s+1)+s, a(v, u')}} = \mathbf{0} 
\end{aligned}
\end{equation}
For all $i = 0, 1, \ldots, r-1$ and all $a \in \{a~:~0 \leq a \leq l'-1, a_v = 0 \}$, all terms in \eqref{eqn:opt-repair-Hc=0-2} apart from the $2s$ underlined terms can be found from the data chunks in $\mathcal{M}^{(v, s)}$. %
For $0 \leq u' \leq s-1$, let $\mathbf{b}_{u'}$ denote $\mathbf{\Psi}_1\sum\nolimits_{w=0}^{u'-1}\mathbf{c}_{v(s+1)+u',w}+ \mathbf{c}_{v(s+1)+u',u'}+\mathbf{\Psi}_2\sum\nolimits_{w=1}^{s-1-u'}\mathbf{c}_{v(s+1)+u',u'+w}+ \mathbf{\Psi}_4\sum\nolimits_{w=1}^{u'}\mathbf{c}_{v(s+1)+u'-w,u'}+ \mathbf{\Psi}_3\sum\nolimits_{w=1}^{s-1-u'}\mathbf{c}_{v(s+1)+u'+w,u'}$. %
The $2s$ underlined terms in \eqref{eqn:opt-repair-Hc=0-2} can be further written as
\begin{equation}
\label{eqn:opt-repair-Hc=0-3}
\begin{aligned}
&\begin{bmatrix}
\sum\nolimits_{u'=0}^{s-1}\mathbf{A}_{0, 2vs+u'}\mathbf{b}_{u'}+ \sum\nolimits_{u' = 0}^{s-1}\mathbf{A}_{0,2(v+1)s-s+u'}\mathbf{c}_{v(s+1)+s, a(v, u')} \\
\sum\nolimits_{u'=0}^{s-1}\mathbf{A}_{1, 2vs+u'}\mathbf{b}_{u'}+ \sum\nolimits_{u' = 0}^{s-1}\mathbf{A}_{1,2(v+1)s-s+u'}\mathbf{c}_{v(s+1)+s, a(v, u')} \\
\vdots \\
\sum\nolimits_{u'=0}^{s-1}\mathbf{A}_{r-1, 2vs+u'}\mathbf{b}_{u'}+ \sum\nolimits_{u' = 0}^{s-1}\mathbf{A}_{r-1,2(v+1)s-s+u'}\mathbf{c}_{v(s+1)+s, a(v, u')} \\
\end{bmatrix} \\
&= 
\begin{bmatrix}
\mathbf{A}_{0,2vs} & \ldots & \mathbf{A}_{0, 2vs+s-1} & \mathbf{A}_{0,2(v+1)s-s} & \ldots & \mathbf{A}_{0,2(v+1)s-1} \\
\mathbf{A}_{1,2vs} & \ldots & \mathbf{A}_{1, 2vs+s-1} & \mathbf{A}_{1,2(v+1)s-s} & \ldots & \mathbf{A}_{1,2(v+1)s-1} \\
\vdots & \vdots & \vdots & \vdots & \vdots & \vdots \\
\mathbf{A}_{r-1,2vs} & \ldots & \mathbf{A}_{r-1, 2vs+s-1} & \mathbf{A}_{r-1,2(v+1)s-s} & \ldots & \mathbf{A}_{r-1,2(v+1)s-1}
\end{bmatrix}
\begin{bmatrix}
\mathbf{b}_0 \\
\vdots \\
\mathbf{b}_{s-1} \\
\mathbf{c}_{v(s+1)+s,a(v,0)} \\
\vdots \\
\mathbf{c}_{v(s+1)+s,a(v,s-1)}
\end{bmatrix}
\end{aligned}
\end{equation}

According to \eqref{eqn:opt-repair-Hc=0-3}, since the $(K+r, K, m)$ base code is MDS, the data chunks $\{ \mathbf{b}_0, \ldots, \mathbf{b}_{s-1},   \mathbf{c}_{v(s+1)+s, a(v,0)}, \ldots, \mathbf{c}_{v(s+1)+s, a(v,s-1)}\}$ can be determined from $\{ \sum\nolimits_{u'=0}^{s-1}\mathbf{A}_{i, 2vs+u'}\mathbf{b}_{u'} + \sum\nolimits_{u' = 0}^{s-1}\mathbf{A}_{i,2(v+1)s-s+u'}\mathbf{c}_{v(s+1)+s, a(v, u')} ~:~0 \leq i \leq r-1\}$ for all $a \in \{a~:~0 \leq a \leq l'-1, a_v = 0\}$. %
Specifically, the data chunks 
\begin{equation*}
\{\mathbf{c}_{v(s+1)+s, a}~:~0 \leq a \leq l'-1\} = \{ \mathbf{c}_{v(s+1)+s, a(v, u')}~:~a_v = 0, 0 \leq u' \leq s-1 \}
\end{equation*}
are uniquely determined by the data chunks 
\begin{equation*}
\{ \sum\nolimits_{u'=0}^{s-1}\mathbf{A}_{i, 2vs+u'}\mathbf{b}_{u'}+ \sum\nolimits_{u' = 0}^{s-1}\mathbf{A}_{i,2(v+1)s-s+u'}\mathbf{c}_{v(s+1)+s, a(v, u')}~:~a_v = 0, 0 \leq i \leq r-1\}. 
\end{equation*}
As mentioned above, the values $\{ \sum\nolimits_{u'=0}^{s-1}\mathbf{A}_{i, 2vs+u'}\mathbf{b}_{u'}+ \sum\nolimits_{u' = 0}^{s-1}\mathbf{A}_{i,2(v+1)s-s+u'}\mathbf{c}_{v(s+1)+s, a(v, u')}~:~a_v = 0, 0 \leq i \leq r-1\}$ are uniquely determined by the data chunks in $\mathcal{M}^{(v, s)}$. %
Consequently, $\{\mathbf{c}_{v(s+1)+s, a}~:~0 \leq a \leq l'-1\}$ are uniquely determined by data chunks in $\mathcal{M}^{(v, s)}$, that is, node $\mathbf{c}_{v(s+1)+s}$ can be recovered from the data chunks in $\mathcal{M}^{(v, s)}$. %
We now conclude that node $\mathbf{c}_{v(s+1)+u}$ can be recovered from the data chunks in $\mathcal{M}^{(v, u)}$. %
The proof is complete. %

\end{document}